# Chemically-Informed Machine Learning Approach for Prediction of Reactivity Ratios in Radical Copolymerization

Habibollah Safari[1], Mona Bavarian[1*]

[1] Department of Chemical and Biomolecular Engineering, University of Nebraska-Lincoln, Lincoln, NE 68588, USA

* Corresponding author

E-mail address: mona.bavarian@unl.edu ; Tel.: +1-402-472-5399

**ABSTRACT**

Predicting monomer reactivity ratios is crucial for controlling monomer sequence distribution in copolymers and their properties. Traditional experimental methods of determining reactivity ratios are time-consuming and resource-intensive, while existing computational methods often struggle with accuracy or scalability. Here, we present a method that combines unsupervised learning with artificial neural networks to predict reactivity ratios in radical copolymerization. By applying spectral clustering to physicochemical features of monomers, we identified three distinct monomer groups with characteristic reactivity patterns. This computationally efficient clustering approach revealed specific monomer group interactions leading to different sequence arrangements—including alternating, random, block, and gradient copolymers—providing chemical insights for initial exploration. Building upon these insights, we trained artificial neural networks to achieve quantitative reactivity ratio predictions. We explored two integration strategies: direct feature concatenation, and cluster-specific training, which demonstrated performance enhancements for targeted chemical domains compared to general training with equivalent sample sizes. However, models utilizing complete datasets outperformed specialized models trained on focused subsets, revealing a fundamental trade-off between chemical specificity and data availability. This work demonstrates that unsupervised learning offers rapid chemical insight for exploratory analysis, while supervised learning provides the accuracy necessary for final design predictions, with optimal strategies depending on data availability and application requirements.

# 1. INTRODUCTION

Homopolymers have demonstrated significant roles across various industries, including microelectronics[1], pharmaceutical manufacturing[2], and separation technology[3]. Building upon this foundation, multi-component copolymerization involving two, three, or more monomers has expanded polymer science. This advancement allows for the precise fine-tuning of physical and chemical properties in copolymers[4,5], terpolymers[6,7], and other complex structures[8], enabling these materials to meet the rigorous challenges and demands of both industrial applications and academic research. Nevertheless, the synthesis and characterization of these materials present significant challenges in polymer science, requiring substantial investments in both time and resources.

The intrinsic relationship between monomer sequence distribution in copolymers and their resultant physical and chemical properties[9] has necessitated the development of predictive frameworks for understanding monomer sequence distribution without extensive experimental work. This predictive strategy can be beneficial through two complementary approaches: The first one known as the forward design problem addresses the fundamental challenge of predicting final monomer sequence distribution prior to synthesis. This predictive capability optimizes the development process by enabling researchers to anticipate material properties before committing resources to synthesis. Complementing this approach, the inverse design problem employs a reverse-engineering methodology, working backward from desired material properties to identify optimal monomer combinations. This method facilitates the targeted development of novel polymers with specific properties, thereby accelerating innovation in polymer science[10]. Combining the two approaches—predicting the final product's arrangement through forward design and deducing optimal monomer structures for targeted outcomes via backward design—leads to deeper understanding of the critical role of monomer sequence prediction in reaction engineering.

The sequence distribution of monomer units represents one of the most critical aspects in determining copolymer chain microstructure and, consequently its final properties. This distribution can be quantitatively characterized through reactivity ratios ($r_{ij}$), a kinetic concept that describes the relative preference of a growing polymer chain to add either type of monomer during copolymerization[11]. Although this concept was initially developed for binary copolymer systems, it has been successfully extended to more complex polymerization systems including terpolymerization and other multicomponent copolymerization[12,13,14,15]. For example, in copolymerization, different values of the reactivity ratios result in distinct sequence distributions, such as alternating, random, block, or gradient arrangements. In the case of a binary system, the reactivity ratios can be expressed as:

$$r_1 = \frac{k_{p,11}}{k_{p,12}}, \quad r_2 = \frac{k_{p,22}}{k_{p,21}} \tag{1}$$

Here, the $k_{p,xy}$ is the propagation rate constant of radical $x$ with monomer species $y$. For instance, poly (styrene-co-maleic anhydride), SMA, is a classic alternating copolymer system, with $r_1$ = 0.001 and $r_2$ = 0.097 [16, 17]. These small values for reactivity ratios result in a strong preference for cross-propagation, producing a perfectly alternating sequence. In contrast, poly (acrylonitrile-co-glycidyl methacrylate) is a representative random copolymer system, where the reactivity ratios $r_1$ = 0.95 and $r_2$ = 0.85 are close to unity [18]. Such values indicate that both monomers have comparable tendencies toward self- and cross-propagation. Poly (N-vinylpyrrolidone-co-N-vinylcaprolactam) represents a block copolymer system, with reported reactivity ratios of $r_1 = 2.8$ and $r_2 = 1.7$ [19]. Gradient copolymers are obtained when the reactivity ratios are highly asymmetric ($r_1 > 1$ and $r_2 < 1$ or vice versa), leading to a continuous drift in composition along the chain. For instance, systems in which methacrylic acid and 2-ethylacrylic acid ($r_1 = 2.29$ and $r_2 = 0.06$) [20] or methyl methacrylate and 2-naphthyl methacrylate ($r_1 = 0.58$ and $r_2 = 2.53$) [21] go through radical copolymerization are examples of gradient copolymers as shown in the figure 1.

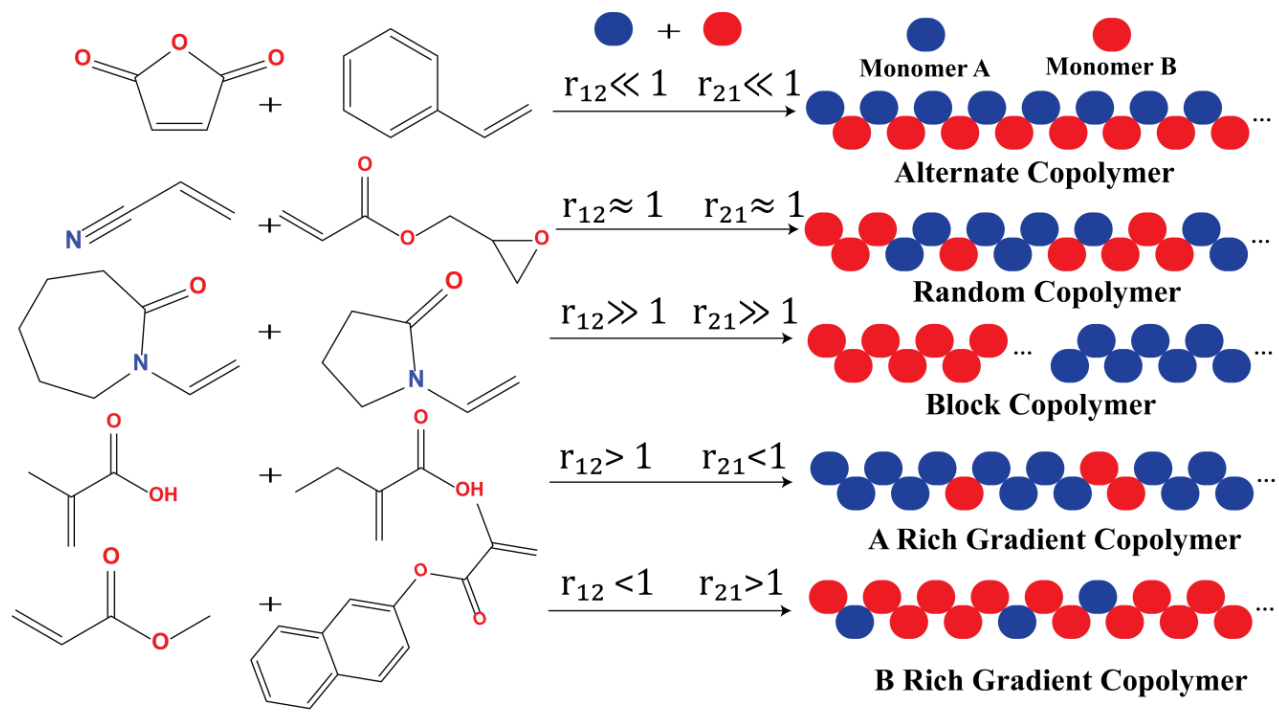

Figure 1. Different reactivity ratios in copolymerization leads to different sequence distribution resulting in various properties.

To this end, leveraging non-empirical approaches for estimation of reactivity ratio values offers significant advantages from a reaction engineering perspective. This need has become increasingly urgent as the field advances toward sophisticated sequence-controlled polymer design through the convergence of computational prediction, experimental characterization, and synthetic control methodologies. Recent breakthroughs in molecular dynamics simulations combined with machine learning frameworks now enable targeted inverse design of polymer sequences with specific conformational properties[22]. While advanced experimental techniques now allow direct sequence analysis of copolymers[23] and practical applications demonstrate how reactivity ratios serve as fundamental levers for synthetic sequence control[24]. Building upon these advances, computational methods for predicting reactivity ratios have become essential tools that bridge the gap between sequence design algorithms and practical synthetic implementation. These predictive approaches can accelerate polymer synthesis and facilitate the discovery of new copolymers by predicting the final product's sequence distribution and properties before experimental synthesis[25]. However, existing computational approaches present distinct challenges[26]. Density Functional Theory (DFT) calculations, while accurate in predicting reactivity ratios[27] and kinetic constants, are computationally intensive and time-consuming, making them impractical for screening large monomer combinations - a challenge known as the scalability problem. Traditional machine learning techniques, including neural networks, offer computational efficiency but often struggle to achieve high accuracy due to the complex nature of polymerization reactions[28,29]. To overcome these limitations, a hybrid framework that uniquely integrates feature engineering with machine learning architectures is presented. In this approach, monomers are systematically clustered based on their physicochemical features, allowing for specialized models to predict reactivity ratios for interactions between distinct clusters. This clustering strategy enhances prediction accuracy by accounting for the inherent chemical similarities and differences between monomer groups.

The organization of this paper is as follows: Section 2 presents our methodology, encompassing data curation and engineering, framework architecture, molecular feature extraction, clustering analysis, and model training. Section 3 provides comprehensive results and discussion, including model performance evaluation and chemical interpretation of predictions. Finally, Section 4 concludes the paper with key findings and future research directions.

## 2. METHODOLOGY

### 2.1 Data Curation and Engineering

For this study, data were collected through utilization of CoPolDB, which is a database for radical copolymerization developed by Takahashi et al.[30]. This database represents molecular names, SMILES (Simplified Molecular Input Line Entry System) notation, and corresponding reactivity ratios. Reactivity ratio in copolymerization has a fundamentally statistical nature [31] and are influenced by both molecular structure and experimental conditions such as temperature and solvent. However, it has been observed that for some monomers, the effect of solvent is not significant [32,33,34]. While temperature seems to have more effect compared to the solvent, it is reported for some systems that doubling the reaction temperature can only increase the reactivity ratio up to 20% [32,35]. Nguyen et al. and Farajzadehahari et al. developed ML models based on the assumption that reactivity ratios depend solely on monomer structure [26,28]. In this study, we assume that reactivity ratios are primarily governed by molecular structure, including steric hindrance, [36] electronic effects (such as electron-donating or electron-withdrawing) [37], and ability of the monomers to stabilize the radical after adding to the growing polymer chain (resonance effect) [38].

The initial dataset from CoPolDB included 2991 reactivity ratio entries, which contained multiple experimental measurements for some monomer pairs. Since reactivity ratios are defined as ratios of kinetic parameters during propagation and must therefore be positive values, we filtered out all entries where one or both reactivity ratios were negative. This initial quality control step removed 296 entries, leaving 2695 entries for further processing. Additionally, monomers containing multiple vinyl groups were excluded from the dataset, as they present unique challenges for determining reactivity ratio. As demonstrated by Wiley and Sale[39] and Storey [40], divinyl monomers effectively create three-component copolymerization systems rather than true binary systems. For instance, Storey reported that p-divinylbenzene copolymerization data could not produce constant reactivity ratios across different monomer compositions, with the variability attributed to concurrent polymerization of pendant vinyl groups [40]. Therefore, to maintain the validity of binary copolymerization assumptions underlying our modeling approach, entries containing monomers with multiple vinyl groups were excluded from subsequent analysis. Stage 2 filtering identified and removed 330 entries containing multi-vinyl monomers, leaving 2365 entries for further processing.

In addition, the dataset contained repeated copolymerization experiments for some paired monomers. To ensure the robustness of the data, we calculated the normalized variance ($\sigma^2/\mu^2$) for both $r_1$ and $r_2$ values across all repeated experiments for each monomer pair. This dimensionless metric provides a consistent measure of experimental variability relative to the mean value, enabling fair comparison across different reactivity ratio scales. Specifically, if the normalized variance for a given monomer pair was less than 0.2, we considered the results sufficiently consistent to proceed with averaging the $r_1$ and $r_2$ values. This approach allowed us to combine redundant data while retaining the integrity of the experimental results. For the subsequent stage, we focused on the monomer pairs that exhibited low normalized variance, ensuring that only reliable data was used for training. Within these selected monomer pairs, we averaged the $r_1$ and $r_2$ values, resulting in a unique reactivity ratio for each copolymerization:

$$D = \left\{ r_{1_{avg}}, r_{2_{avg}} \mid \frac{\sigma_{r_1}^2}{\mu_{r_1}^2} \leq 0.2, \frac{\sigma_{r_2}^2}{\mu_{r_2}^2} \leq 0.2 \right\} \tag{3}$$

$$r_{1\_avg} = \frac{\sum r_{1,i}}{n}, r_{2\_avg} = \frac{\sum r_{2,i}}{n} \tag{4}$$

This filtering process analyzed 376 monomer pairs with multiple measurements, of which 169 pairs met the consistency criteria and were retained with averaged values, while 207 pairs were excluded due to excessive experimental variability. Combined with 1,175 pairs having single measurements, this stage yielded a final dataset of 1,344 reliable unique monomer pairs.

To maximize the utilization of available experimental data, we augmented the dataset by leveraging the inherent symmetry in copolymerization reactivity. In radical copolymerization, the reactivity relationship between two monomers is fundamentally symmetric: if monomer pair (A, B) exhibits reactivity ratios ($r_1$, $r_2$), then the reversed pair (B, A) will exhibit reactivity ratios ($r_2$, $r_1$) due to the reciprocal nature of the kinetic relationships. Each monomer pair was systematically duplicated with complete symmetry, where monomer identities, SMILES representations, and reactivity ratios were correspondingly swapped. This augmentation strategy doubled the dataset size from 1,344 to 2,688 entries while preserving the underlying chemical relationships and introducing no artificial bias.

To further ensure data quality and model reliability, we applied bounds of $0.01 \leq r \leq 10$ for both reactivity ratios. This range excludes extreme values that often represent experimental outliers or systems where the terminal model assumptions break down, while encompassing the vast majority of commercially relevant copolymerization systems. Very low reactivity ratios ($r < 0.01$) were excluded because they represent extreme cross-propagation behavior where reactivity ratio determination becomes highly sensitive to experimental uncertainties in monomer feed composition and conversion measurements, leading to poor data reliability that could negatively impact machine learning model performance. Very high values ($r > 10$) indicating near-complete homopolymerization preference were excluded as they typically involve specialized systems with poor experimental reproducibility. This filtering process removed 384 entries from the dataset, yielding a final dataset of 2,304 high-quality measurements fully compatible with logarithmic transformation.

Figure 2 illustrates the final curated dataset characteristics without symmetry-based augmentation. Panel A shows the distribution of 1152 unique experimental reactivity ratio pairs in $\log_{10}$ space, clearly delineating different copolymerization domains: alternating tendency, random tendency, block tendencies, and gradient regions. Panels B and C demonstrate that log transformation effectively converted the highly skewed original distributions (3.10 and 3.40) into near-Gaussian distributions (-0.48 and -0.37), providing an optimal foundation for machine learning model development.

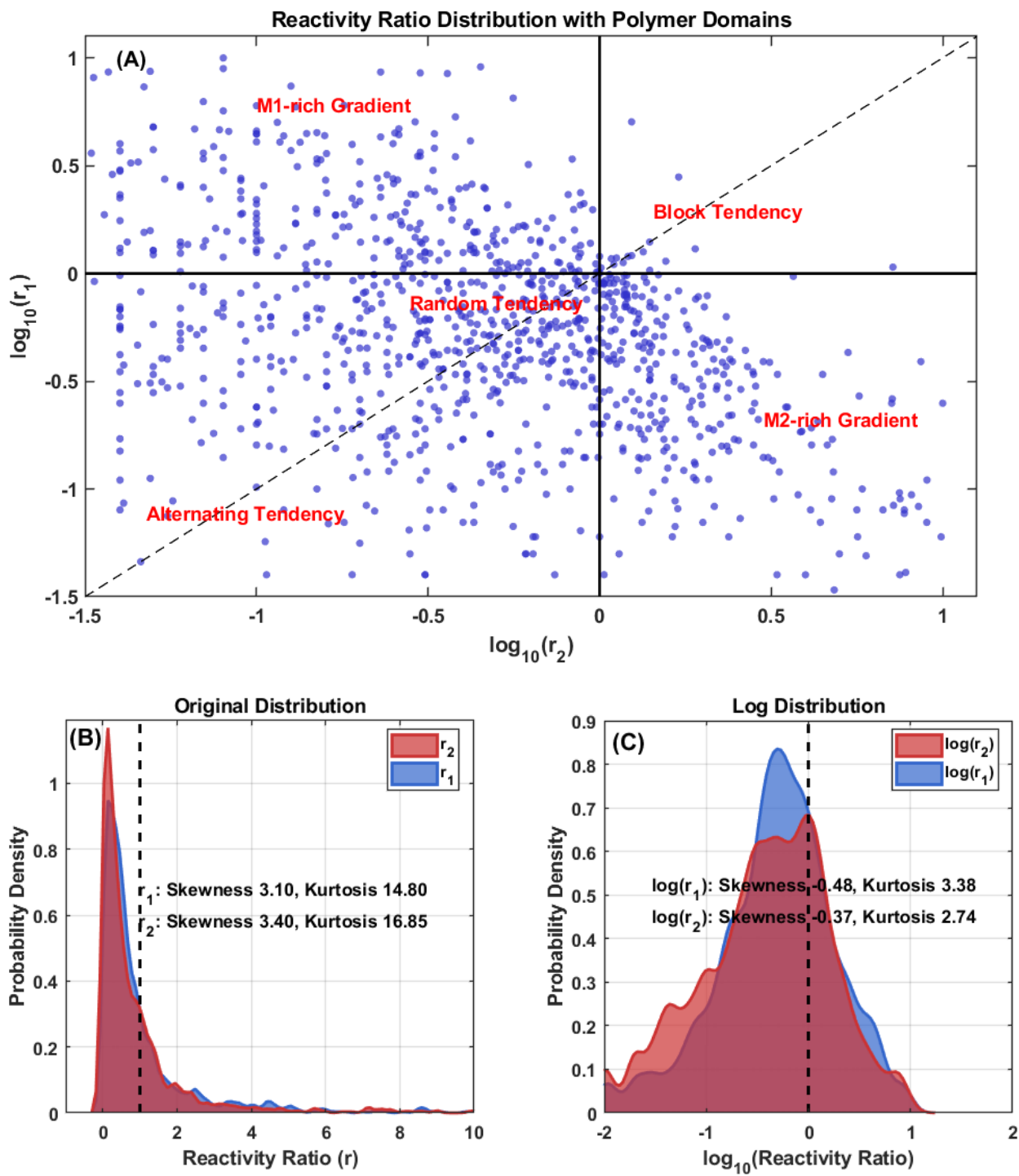

Figure 2. Curated reactivity ratio dataset: (A) Distribution in log₁₀ space showing copolymerization domains, (B) Original skewed distributions, and (C) Log-transformed near-Gaussian distributions.

## 2.2 Framework & Feature Extraction

The proposed framework introduces a chemically informed machine learning approach for predicting reactivity ratios in copolymerization systems. First, the important physicochemical features of monomers are extracted from their molecular structures. Then, based on optimal clustering algorithms and cluster numbers, monomers are divided into different groups, with each group exhibiting characteristic reactivity patterns. This clustering reveals specific monomer group interactions that correspond to different copolymer sequence arrangements. Subsequently, two distinct integration strategies are evaluated to investigate how clustering insights can affect machine learning performance. Figure 3 illustrates the systematic development stages of our proposed framework for reactivity ratio prediction in copolymerization systems.

Figure 3. Multi-model AI framework for predicting copolymer entity distribution: Molecular feature extraction leads to monomer clustering,

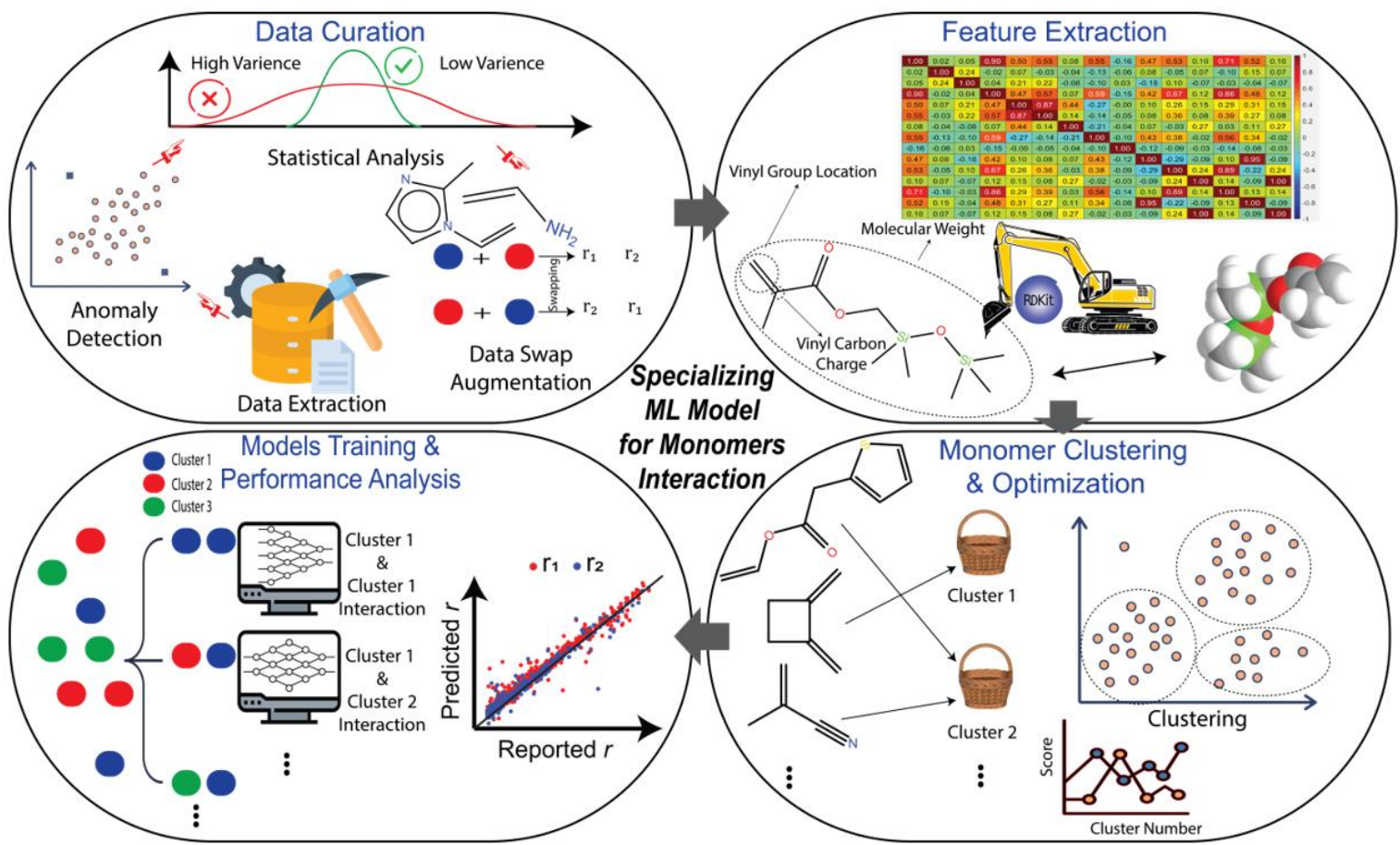

enabling chemical-informed machine learning models for reactivity ratio prediction.

Here, after clustering of monomers in k groups based on their physicochemical attributes, the number of unique monomer interactions can be calculated from this equation:

$$M = \binom{k+n-1}{n} = \frac{(k+n-1)!}{n!\,(k-1)!} \tag{5}$$

where M is the number of monomer interactions, k is the number of clusters to which monomers belong, and n is defined based on the number of monomers participating in multicomponent polymerization (n = 2 in copolymerization, n = 3 in terpolymerization, etc.). This formula accounts for all possible combinations of cluster interactions, including cases where monomers come from the same cluster. For instance, in a copolymerization system (n=2) with three clusters (k=3), six different types of interactions between monomers can occur during polymerization ((3+2-1)! / (2! (3-1)!) = 4! / (2!2!) = 6). These six interactions represent all possible combinations of how monomers from each cluster can react during polymerization. For example, a monomer from Cluster 1 can react with another monomer from Cluster 1 (self-interaction), or it can react with a monomer from Cluster 2 or Cluster 3 (cross-interactions).

For clustering monomers and determining the number of groups to which monomers belong, firstly fifteen properties from each monomer were extracted. This selection of molecular descriptors was guided by both expert knowledge and rigorous statistical analysis of structure-reactivity relationships in our dataset. For instance, as shown in Figure 4, analyzing the position of vinyl groups revealed a striking difference in reactivity ratios: monomers with vinyl groups not attached to a ring showed significantly higher reactivity (mean = 1.35) compared to those with ringed vinyl groups (mean = 0.67). This observation aligns with mechanistic understanding, as vinyl groups embedded in ring structures face greater steric constraints, reducing their accessibility during propagation. Moreover, monomers with vinyl groups in cyclic structures require high energy for ring-breaking during propagation. These observations can be interpreted from the statistical analysis of this structural property. This structural impact is further illustrated by comparing seemingly similar monomers like styrene (mean = 1.50) and maleic anhydride (mean = 0.16), where differences in vinyl group accessibility and electronic environment lead to dramatically different reactivity patterns. These analyses strongly support the inclusion of vinyl group characteristics as key features for monomer clustering in our predictive framework.

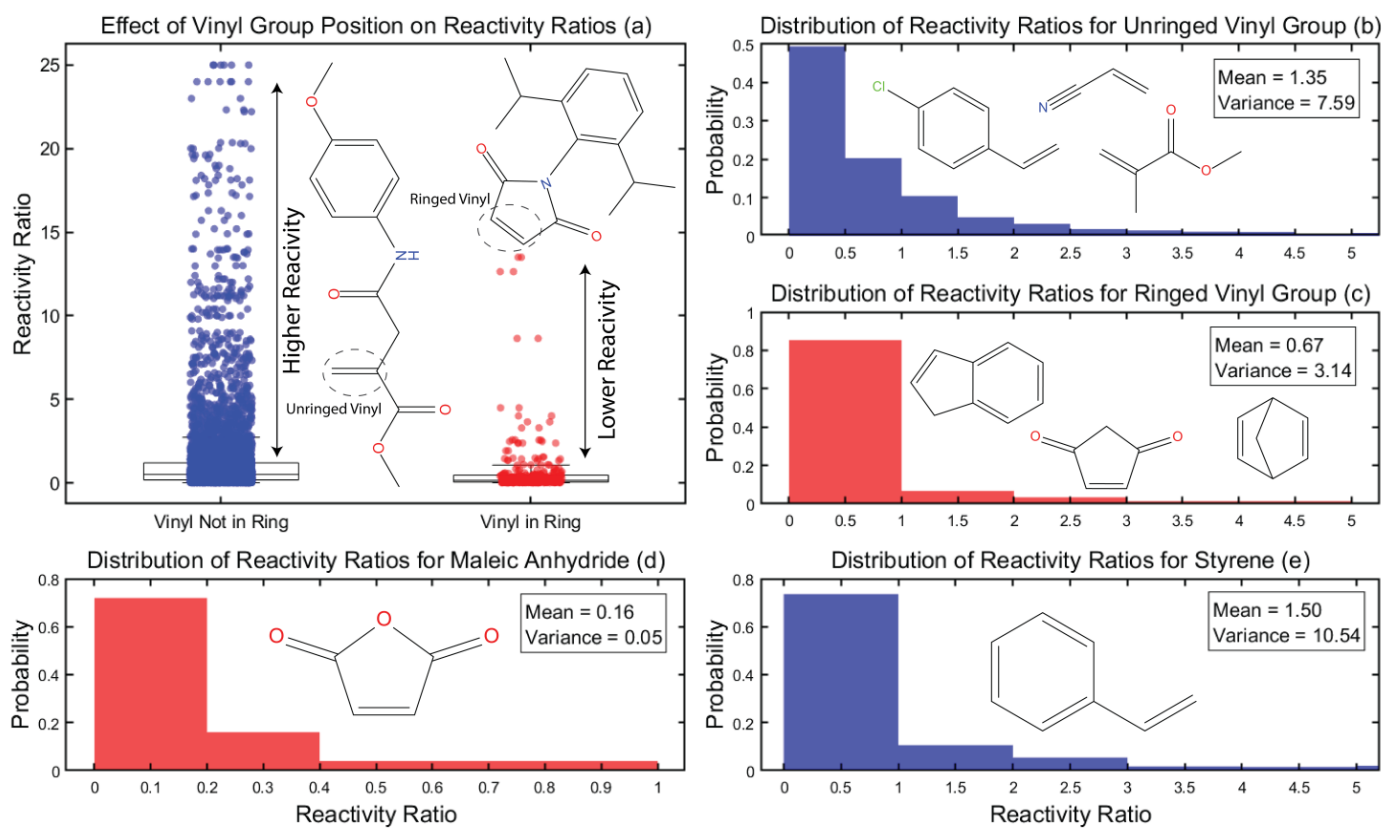

Figure 4. Analysis of vinyl group location impact on reactivity of monomers : (a) Impact of vinyl group position, comparing ringed vs unringed configurations (b) Reactivity ratio distribution for vinyl groups not attached to a ring (mean = 1.35) (c) Reactivity ratio distribution for ringed vinyl groups (mean = 0.67) (d) Reactivity ratio distribution for maleic anhydride (mean = 0.16) (e) Reactivity ratio distribution for styrene (mean = 1.50), showing distinct reactivity patterns despite similarities in size and molecular weight.

Building on these statistical insights and knowledge of domain, we expanded our feature set to capture other key parameters affecting radical copolymerization mechanisms. Primary features include electronic properties (atomic charges in vinyl groups), molecular characteristics (volume and lipophilicity), and structural elements (hybridization states, hydrogen bonding characteristics). These parameters were selected for their demonstrated influence on radical stability, steric effects, and monomer-monomer interactions. Additional descriptors like molecular weight, conjugation patterns, and stereochemistry were incorporated to account for spatial effects and approach geometry during polymerization. This approach to feature selection, validated by both statistical analysis and mechanistic understanding, provides a robust foundation for distinguishing different polymerization behaviors. Table 1 summarizes these features and their chemical rationale for monomer clustering in our framework.

Table 1. Molecular Descriptors and Their Chemical Rationale for Monomer Clustering

| Properties | Chemical Rationality for Consideration | Properties | Chemical Rationality for Consideration |
|---|---|---|---|
| Vinyl is Cyclic or Linear | Pi Bond Stability/Accessibility | Molecular LogP | Solubility Effects/Phase Behavior |
| Vinyl Carbons Charge | Electronic Effects on Radical Stability | Hybridization sp/ sp$^2$/sp$^3$ | Electronic Distribution/Orbital Overlap |
| Molecular Volume | Spatial Effect in Propagation/Steric Hindrance | Chirality | Stereochemical Effects on Approach |
| Molecular Weight | Size Effect on Diffusion/Mobility | Num Connected H | Chain Transfer Potential/Reactivity Sites |
| Total Polar Surface Area | Intermolecular Interactions/Polarity Effects | Conjugated Bonds | Resonance Stabilization/Radical Stability |
| Number of H Acceptors | Potential Chain Transfer Sites/H-Bonding Effects | Stereochemistry | Spatial Arrangement/Accessibility Effects |
| Number of H Donors | H-Bonding Capability/Chain Transfer Potential | | |

For effective clustering, the selection of independent molecular features is crucial to avoid model bias. When highly correlated features are used simultaneously in a clustering algorithm, the model tends to cluster monomers based on redundant information, as repeated features do not add new insights to the analysis. This issue, known as the covariance or multicollinearity problem, is a common challenge in machine learning[41]. To address this challenge, we conducted a comprehensive correlation analysis of molecular properties using a correlation heatmap (Figure 5) to identify and eliminate redundant features.

The correlation analysis revealed several closely related molecular features that would not contribute unique information to the model. Examining the correlation heatmap of monomer features, we identified strong correlations between certain properties - notably molecular volume and molecular weight (0.91), as well as sp² hybridization with number of conjugated bonds - while many other features maintained low correlation coefficients, confirming their independence. This statistical analysis, combined with chemical intuition, guided our feature selection to capture distinct aspects of molecular structure and reactivity. By identifying and eliminating highly correlated descriptors, we minimized the multicollinearity problem that could reduce clustering reliability. For example, given the strong correlation between sp² hybridization and conjugated bonds, we selected only one of these features, as they effectively capture the same electronic characteristics. To further validate our feature selection approach, Principal Component Analysis (PCA) was performed on both the original fifteen features and our selected eight features. The analysis revealed significant redundancy in the original feature set, where eight principal components were sufficient to explain 95% of variance, indicating dependencies among the fifteen features. In contrast, our selected eight features required seven components to reach the same threshold, demonstrating that our chosen descriptors maintain high independence with minimal redundancy (Figure S1). A detailed analysis of feature correlations and selection criteria, is provided in the Supporting Information (SI), Section 1.

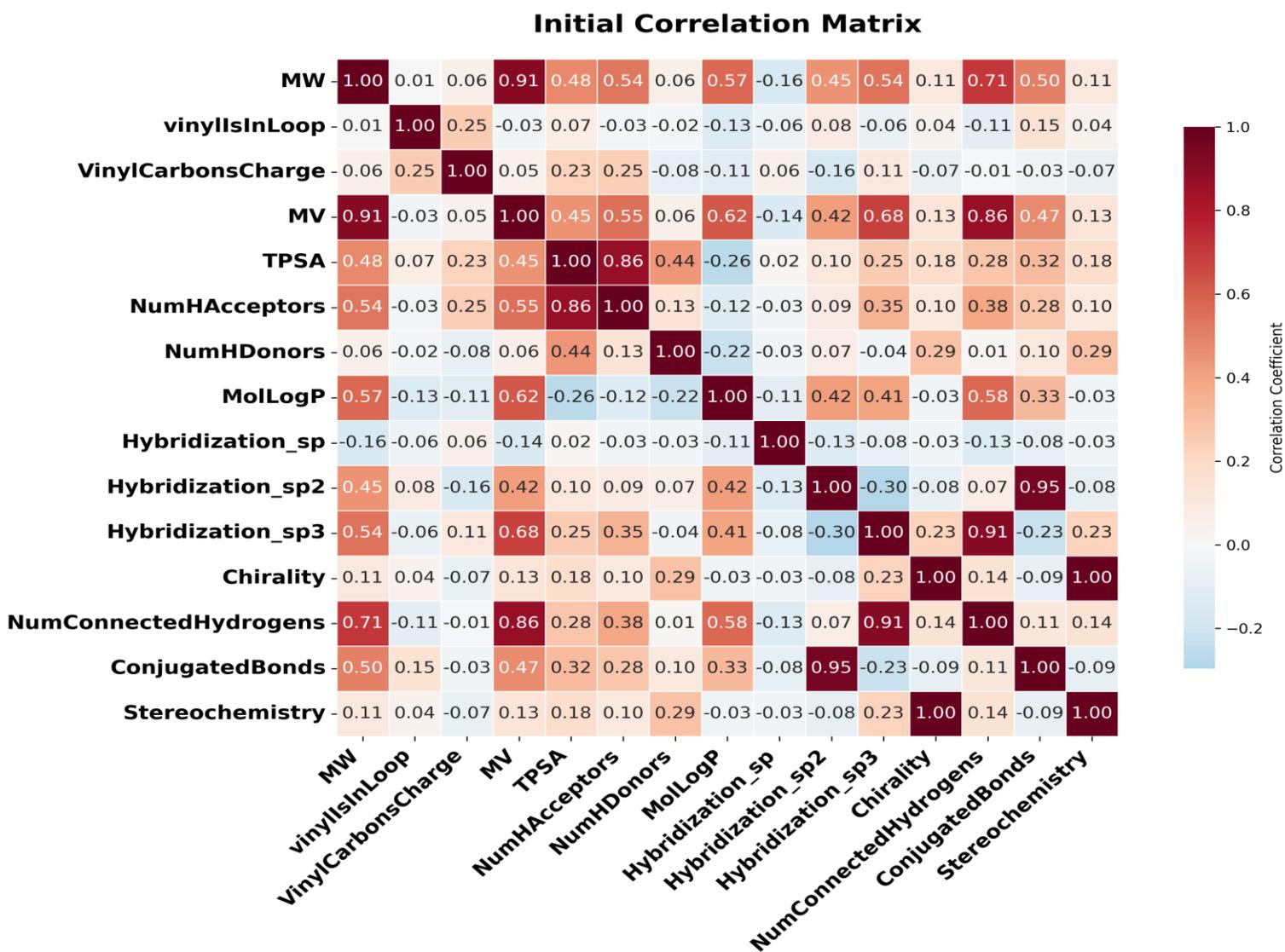

Figure 5. Initial correlation analysis of molecular descriptors in monomer dataset: Heatmap visualization of correlations between fifteen molecular features, revealing key relationships like molecular weight and molecular volume (0.91), sp² hybridization and conjugated bonds (0.95), and total polar surface area with number of H acceptors (0.86), guiding feature reduction for clustering analysis.

## 2.3 Monomer Clustering & Interaction Analysis

Following feature selection, using selected features, multiple clustering algorithms were evaluated across different numbers of clusters to optimize clustering performance. The evaluation encompassed five different algorithms: k-means[42], Spectral Clustering[43], Agglomerative Clustering[44], Gaussian Mixture Model (GMM)[45], and BIRCH[46]. Each algorithm was tested with cluster numbers ranging from 1 to 10. The clustering quality was assessed using three complementary evaluation metrics: silhouette score[47], which measures how similar monomers are to their own cluster compared to other clusters (Figure S2a); Calinski-Harabasz index[48], which evaluates cluster separation by comparing between-cluster to within-cluster variance (Figure S2b); and Davies-Bouldin index[49], which quantifies the average similarity between each cluster and its most similar neighbor (Figure S2c). Based on these metrics, spectral clustering with three clusters emerged as the optimal choice, demonstrating consistently high silhouette scores (0.95-1.0) and uniform performance across all evaluation criteria, indicating well-defined and stable monomer groupings. Detailed information about comprehensive performance comparisons is provided in the Supporting Information (SI), Section 2.

After selecting spectral clustering, the algorithm was implemented on our dataset with $X = \{x_1, ..., x_{628}\}$ of 628 unique monomers with 8 molecular descriptors. A similarity matrix W was constructed using k-nearest neighbors' approach. In this matrix, each element $W_{ij}$ represents the molecular similarity between monomer i and monomer j, calculated using Euclidean distances between their molecular descriptors in the 8-dimensional feature space. For each monomer i, its k nearest neighbors were identified based on these distances, and the corresponding $W_{ij}$ values were set to 1, while all other elements were set to 0.

This similarity matrix was then transformed into a graph Laplacian matrix L, which represents the graph structure of our data and captures both local and global relationships between monomers. The normalized Laplacian matrix was computed as:

$$L = I - D^{-\frac{1}{2}} W D^{-\frac{1}{2}} \qquad (6)$$

The normalized Laplacian matrix L was computed in equation 6, where the term I represents the identity matrix and D denotes the degree matrix with diagonal elements $D_{ii} = \sum_j W_{ij}$, representing the number of neighbors for each monomer. From this Laplacian matrix, the first k eigenvectors were extracted to form a new feature space for clustering, where k=3 was determined as optimal through systematic evaluation of clustering metrics. Detailed information can be found in other reference[43].

Figure 6 presents a comprehensive clustering analysis of monomers based on eight molecular features. Panel A shows the correlation heatmap of the selected features, which are relatively independent and encompass different monomer properties that significantly influence reactivity during the propagation stage of copolymerization. The dimensionality reduction analysis (Supporting Information (SI), Figure S3) was performed using various algorithms (PCA[50], t-SNE[51], and UMAP[52]) to transform the 8-dimensional data into a more interpretable 2D representation. Panel B displays the t-SNE visualization, which reveals three distinct monomer clusters in two-dimensional space. The proximity of points in the t-SNE plot indicates molecular similarity, where closely clustered points represent monomers with similar characteristics.

Panels C and D analyze the cluster characteristics through complementary visualizations (bar plot and radar chart), revealing distinct molecular profiles for each group. Cluster 3 (green) comprises monomers characterized by vinyl groups within rings, high vinyl carbon charge, elevated total polar surface area (TPSA), and fewer hydrogen donors. In contrast, Cluster 2 (blue) is distinguished by monomers with high sp² hybridization counts, larger molecular weights, and increased lipophilicity (MolLogP), while Cluster 1 (red) exhibits intermediate characteristics between these two extremes. These cluster-specific features are particularly evident in the radar chart visualization in Panel D.

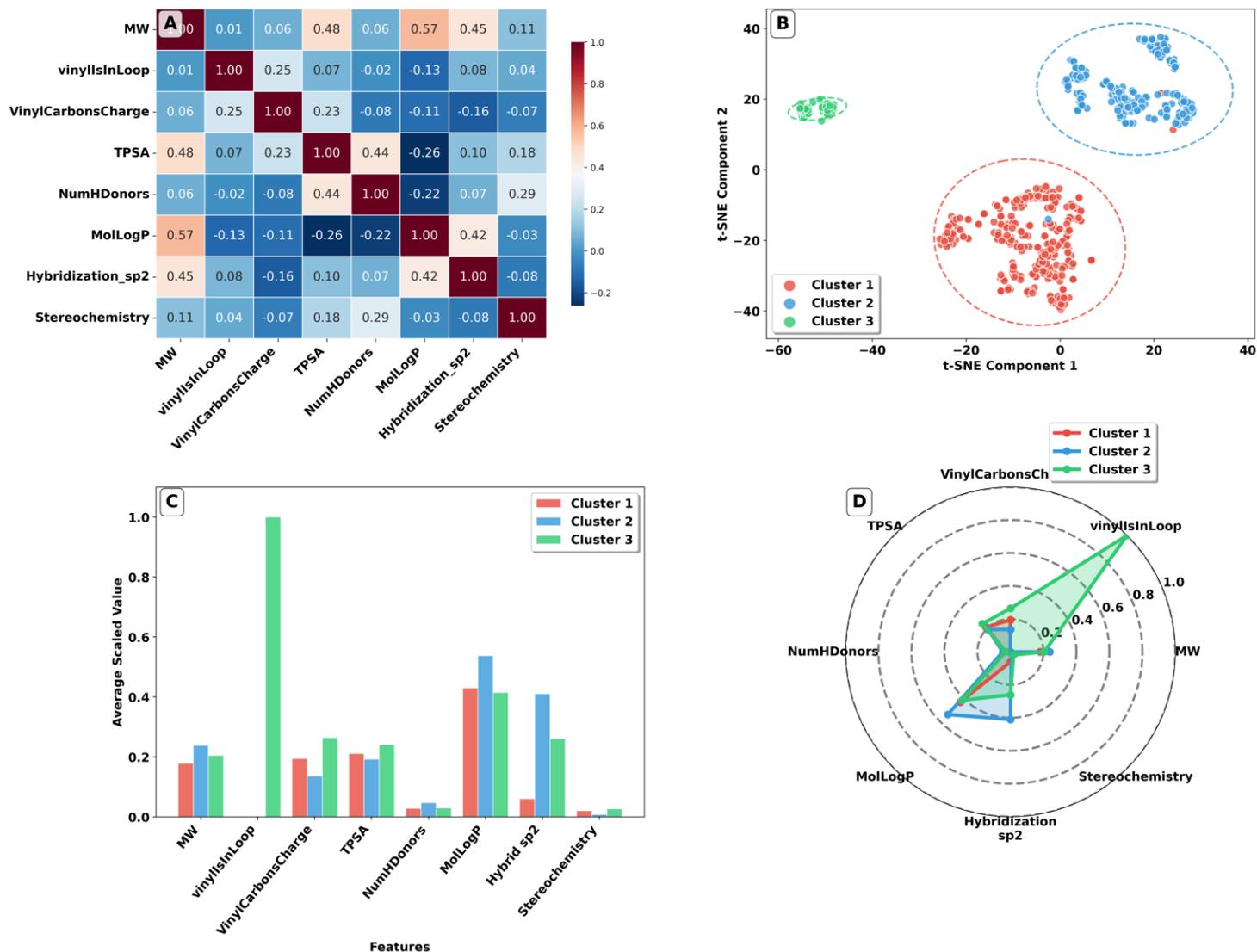

Figure 6. Spectral clustering analysis of monomers using selected molecular descriptors: (A) Correlation heatmap of the eight selected features demonstrating their independence (B) t-SNE visualization of three distinct monomer clusters from spectral clustering algorithm (C) Bar chart showing average feature values across clusters, highlighting key differentiating characteristics (D) Radar chart representation of average feature values, emphasizing distinct molecular property patterns in each cluster.

After assigning corresponding clusters to each monomer in the copolymerization dataset (Cluster 1 with 343 monomers, Cluster 2 with 234 monomers, Cluster 3 including 51 monomers), the reactivity behavior for six distinct interaction groups was analyzed. This statistical analysis (Figure 7) within and between clusters reveals distinct patterns in monomer interactions that provide insights into the likely sequence arrangements in the resulting copolymer chains, including alternating, random, block, and gradient configurations. Specifically, the dataset contained 870 entries for Cluster 1–1 interaction, 858 entries for Cluster 1–2, 154 entries for Cluster 1–3, 360 entries for Cluster 2–2, 52 entries for Cluster 2–3, and 10 entries for Cluster 3–3.

Cluster 1-Cluster 1 interactions demonstrate mean reactivity ratios of approximately 1.1 for both $r_1$ and $r_2$ with substantial standard deviation, as evidenced by the large standard deviation bars. These reactivity ratios near unity, combined with high standard deviation, suggest that C1-C1 interactions can produce a range of sequence arrangements from random to block copolymer configurations depending on the specific monomer pair involved. The significant standard deviation indicates diverse behaviors within this cluster, emphasizing that precise reactivity ratio determination requires individual model predictions rather than relying solely on cluster averages. Cluster 1-Cluster 2 interactions exhibit asymmetric reactivity patterns with moderate $r_1$ values (~0.6) and higher $r_2$ values (~1.1), indicating a tendency toward gradient copolymer formation where the sequence composition changes continuously along the chain. Similarly, Cluster 1-Cluster 3 interactions show distinctly asymmetric behavior with higher $r_1$ values (~1.4) and lower $r_2$ values (~0.7), again suggesting gradient sequence development but with opposite compositional drift. Cluster 2-Cluster 2 interactions display moderate and relatively balanced reactivity ratios (~0.8 for both) with moderate standard deviation, suggesting random to weakly gradient sequence formation. In contrast, Cluster 2-Cluster 3 interactions exhibit markedly different behavior, with extremely low reactivity ratios for both cluster types (~0.5 and ~0.1), indicating alternating copolymer formation with highly predictable sequence arrangements, as evidenced by the minimal standard deviation bars. Cluster 3-Cluster 3 interactions demonstrate moderate reactivity ratios (~0.6 for both) with considerable standard deviation, suggesting predominantly random copolymer formation with potential for gradient scheme depending on the specific monomer pair involved.

While this statistical analysis provides valuable insights into general trends across cluster interactions, it is important to emphasize that these represent a general behavior within each interaction type. The substantial standard deviation observed in several interaction categories, particularly C1-C1 and C3-C3, shows the necessity of developing regression models capable of predicting precise reactivity ratio values for individual monomer pairs. Detail statistical values are presented in Table S1 Supporting Information (SI), Section 4.

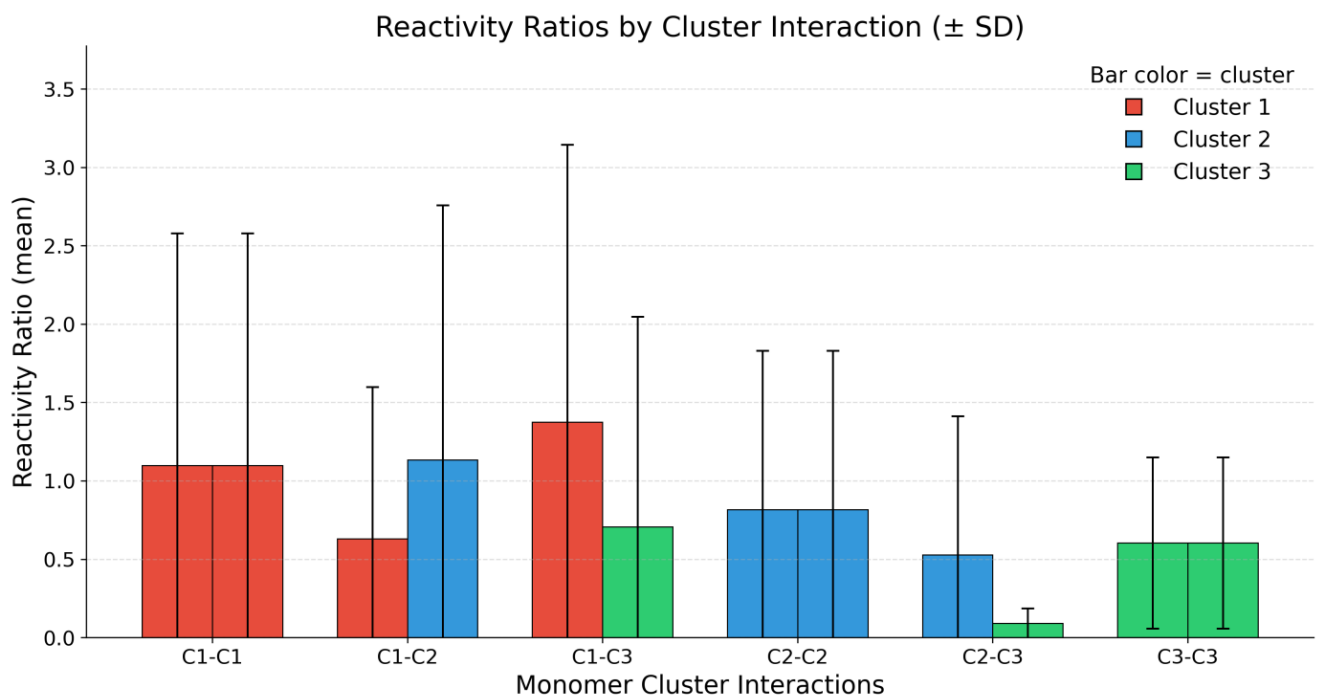

Figure 7. Reactivity ratios pattern for different monomer cluster interactions. Each pair of bars represents the reactivity ratios of participating clusters, where C1, C2, and C3 denote the cluster numbers.

**2.4 Model Training**

In the final stage, we evaluated the impact of incorporating cluster information into machine learning models to investigate whether clustering insights can enhance prediction performance for reactivity ratios. To comprehensively assess this integration, we designed two distinct approaches that independently examine different mechanisms through which clustering information might influence machine learning performance. The first approach employed Direct Feature Integration with Incremental Analysis, while the second approach utilized Controlled Cluster-Specific Training Evaluation. Both approaches were built upon the optimum Fully Connected Neural Network (FCNN) architecture in literature ensuring direct comparability with established methodologies [28].

For model development across both approaches, monomer structures were converted to machine-readable format using Morgan fingerprints[53], an extended circular fingerprint (ECFP) algorithm that transforms molecular structures into bit vectors[54]. The Morgan algorithm systematically assigns initial invariants to each atom based on fundamental atomic properties including atomic number, degree, and valence, then iteratively updates these values by incorporating structural information from neighboring atoms at progressively increasing radii. This iterative process generates unique molecular substructure identifiers that are subsequently hashed into fixed-length bit vectors. Our implementation employed a radius of 3 and generated 2048-bit vectors for each monomer, parameters that have been demonstrated to effectively capture both local and extended structural features while maintaining computational efficiency for large-scale molecular datasets.

In training the model, we considered the logarithms of the reactivity ratios $r_1$ and $r_2$ as the model's outputs, addressing key methodological considerations in reactivity ratio prediction. The log transformation offers several critical advantages: First, since reactivity ratios are intrinsically the ratios of two rate constants (e.g., $r_1 = k_{11}/k_{12}$), transforming them to the log scale converts the multiplicative relationship into an additive one ($\log r_1 = \log k_{11} - \log k_{12}$), which aligns better with the linear combination structure of artificial neural networks that operate primarily through summations of neuron outputs. Second, the log transformation provides balanced error treatment across different reactivity ratio magnitudes—ensuring that prediction errors for small ratios ($r \ll 1$) and large ratios ($r \gg 1$) are weighted equally. Third, the log transformation dramatically improves data normality, reducing skewness, as evidenced before in Figure 2B and 2C.

**2.4.1 Direct Feature Integration**

First approach systematically investigated whether incorporating cluster membership and physicochemical descriptors as explicit input features alongside Morgan fingerprints could enhance predictive performance for reactivity ratio estimation. The methodology employed an incremental analysis framework designed to isolate and quantify the individual contribution of each feature type while maintaining identical neural network architectures and training procedures across all comparisons.

The experimental design implemented three progressively enhanced models to enable precise measurement of feature impact. The baseline model utilized only Morgan fingerprints, concatenating 2048-bit vectors from both monomers to create 4096-dimensional input vectors. The cluster-enhanced model incorporated cluster membership information through one-hot encoding, where each monomer's cluster assignment was transformed into a categorical vector and concatenated with its corresponding Morgan fingerprint. For each monomer pair, the final input vector combined [Morgan A + Cluster A] features with [Morgan B + Cluster B] features. The comprehensive model further extended this approach by integrating eight carefully selected physicochemical descriptors alongside both Morgan fingerprints and cluster information. These molecular properties, were standardized and concatenated in a monomer-specific manner to create the most complete feature representation: [Morgan A + Cluster A + Phys A] combined with [Morgan B + Cluster B + Phys B]. Figure 8 represent the MLP architecture for prediction of reactivity ratio.

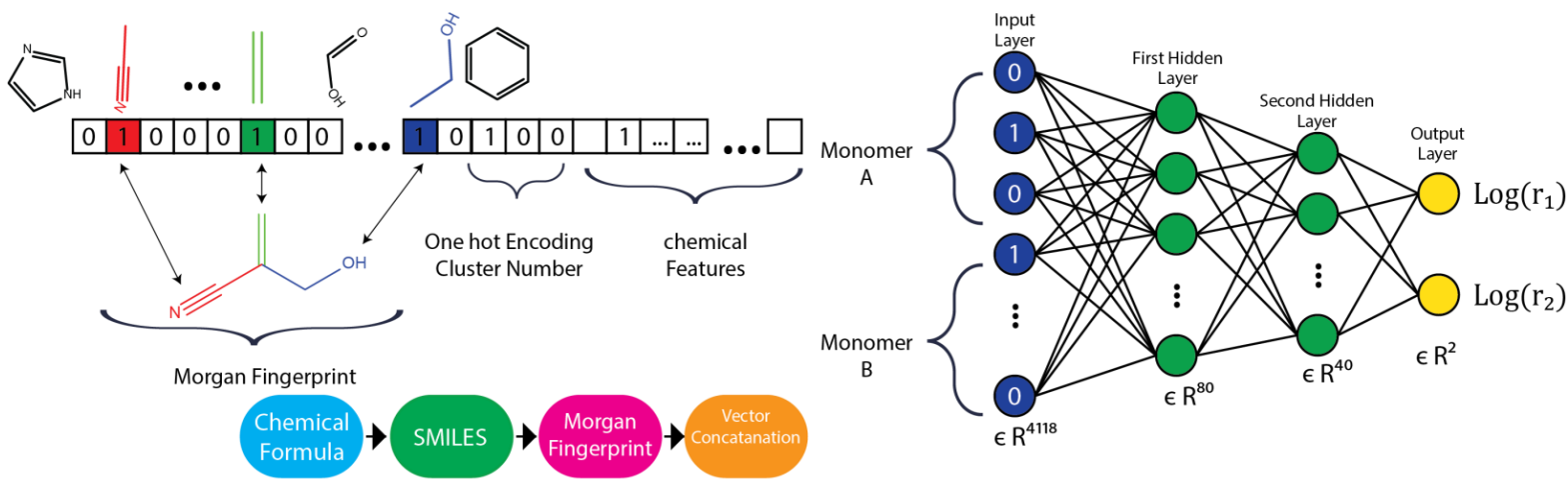

Figure 8. Neural network architecture for direct feature integration approach in reactivity ratio prediction.

Cross-validation analysis was conducted using 10-fold cross-validation to ensure robust statistical evaluation, following established protocols for machine learning validation in chemical property prediction. For each fold, 90% of the data served as the training set while 10% was reserved for testing, with early stopping implemented using 10% of the training data as a validation set to prevent overfitting.

Neural network training employed identical hyperparameters across all three models to ensure fair comparison. The architecture consisted of two hidden layers with 80 and 40 neurons respectively, using ReLU activation functions and linear output layers for log-transformed reactivity ratio prediction. Training utilized the Adam optimizer with a learning rate of 0.001, batch size of 32[28], and mean squared error loss function. Early stopping was implemented with a patience of 30 epochs, monitoring validation loss to restore the best weights and prevent overfitting. This experimental design ensured that any performance differences could be attributed solely to feature composition rather than architectural or training variations.

**2.4.2 Cluster-Specific Training Effect**

To evaluate whether cluster-specific training provides performance advantages over general training approaches, we implemented a controlled experimental framework comparing models trained exclusively on specific cluster interactions against models trained on randomly selected datasets of identical size. The clustering analysis of our dataset revealed six distinct cluster interaction groups with varying data availability: Group 1-1 with 870 entries, Group 1-2 with 858 entries, Group 1-3 with 154 entries, Group 2-2 with 360 entries, Group 2-3 with 52 entries, and Group 3-3 with 10 entries. The fundamental requirement for training a robust FCNN necessitates sufficient training samples to ensure proper model convergence and reliable performance evaluation. Deep neural networks require adequate data volumes to learn complex nonlinear relationships while avoiding overfitting.

Based on these data availability considerations and established guidelines for neural network training, we selected the two largest interaction groups for our controlled comparison study: Cluster 1-1 interactions (870 entries) and Cluster 1-2 interactions (858 entries). These groups provided sufficient samples to support meaningful 10-fold cross-validation while maintaining adequate training set sizes in each fold to ensure reliable model learning. Further investigation of the remaining cluster interactions would require substantially larger datasets, as the current limitation stems directly from data availability constraints that prevent robust neural network training on the smaller interaction groups.

The experimental design implements a controlled comparison between two distinct training methodologies to evaluate the predictive value of cluster-specific training for reactivity ratio estimation. This investigation examines whether models trained exclusively on specific cluster interactions demonstrate superior performance compared to models trained on randomly sampled datasets of equivalent size. For each target cluster interaction, we employed 10-fold cross-validation with two parallel training approaches evaluated on identical test sets. The cluster-specific training methodology utilizes samples exclusively from the target interaction type during model development. When predicting Cluster 1-1 interactions, for instance, the training dataset comprises only monomer pairs where both components belong to Cluster 1, enabling the model to learn patterns specific to this chemical interaction category. The general training methodology serves as the control condition, employing randomly selected samples from the complete dataset while maintaining identical training set sizes to ensure fair comparison. This approach eliminates training set size as a confounding variable, allowing direct assessment of whether chemical similarity within training data contributes to improved predictive performance. The random sampling procedure excludes test set samples to prevent data leakage and maintain experimental integrity.

In both approaches, the training process was performed using 16 CPU cores and 60 GB of RAM on a computational cluster (Holland Computing Center at Nebraska) and conducted using JupyterLab (version 3.4.4.0). The model performance was evaluated using mean square error (MSE) during the training process, while the coefficient of determination ($R^2$) was employed to assess the model's predictive capability on test data:

$$\text{MSE} = \frac{1}{n}\sum_{i=1}^{n}(y_i - \hat{y}_i)^2 \qquad (7)$$

$$R^2 = 1 - \frac{\sum_{i=1}^{n}(y_i - \hat{y}_i)^2}{\sum_{i=1}^{n}(y_i - \bar{y}_i)^2} \tag{8}$$

## 3. RESULT & DISCUSSION

The cross-validation analysis of direct feature integration revealed that incorporating clustering information and physicochemical descriptors as explicit input features provided minimal predictive improvement over the Morgan fingerprint baseline. The baseline model achieved $R^2$ values of 0.4435 ± 0.0610 for $r_1$ and 0.4782 ± 0.0751 for $r_2$ predictions. These results demonstrate improved performance compared to literature[25], where the test set MSE values of 2.31-3.11 using similar Morgan fingerprint neural networks were reported, while our approach achieved MSE values of ~0.19 in log scale (Test $MSE_{r_1} = 0.2043 \pm 0.0279$, Test $MSE_{r_2} = 0.1905 \pm 0.0349$) and ~1.4 actual ($MSE_{r_1} = 1.5061 \pm 0.4753$, $MSE_{r_2} = 1.3941 \pm 0.4064$). The enhanced performance likely stems from our systematic data curation protocols that removed problematic entries and improved logarithmic transformation strategies. Adding cluster membership information through one-hot encoding resulted in virtually unchanged performance with MSE values of 0.1984 ± 0.0142 for $r_1$ and 0.1958 ± 0.0343 for $r_2$. The comprehensive model incorporating all available features showed only marginal improvement, achieving MSE values of 0.1949 ± 0.0187 for $r_1$ and 0.1863 ± 0.0306 for $r_2$ (Figure 9 A, B). Comprehensive comparison between performance results of three models can be found in Table S2 Supporting Information (SI), Section 5.

This limited improvement despite the demonstrated chemical relevance of clustering information can be attributed to a fundamental dimensionality imbalance inherent in the concatenation strategy. The Morgan fingerprints contribute 4,096 features per monomer pair, representing the overwhelming majority of the input space, while cluster encoding adds only 6 features and physicochemical descriptors contribute 16 additional features. This dramatic disparity means that clustering and physicochemical information collectively represents about 0.5% of the total feature space. In neural network training, features with higher dimensionality typically dominate the learning process through their greater influence on gradient updates during backpropagation. Consequently, the chemically meaningful but low-dimensional clustering information becomes effectively invisible to the optimization algorithm.

These results reveal a critical limitation of direct concatenation approaches when attempting to integrate chemical insights with high-dimensional molecular representations. The failure to observe meaningful improvement does not indicate that clustering information lacks predictive value, but rather demonstrates that simple feature concatenation cannot effectively leverage the chemical organization captured through unsupervised learning. This finding necessitates alternative integration strategies that can more effectively harness cluster-based chemical insights, motivating the subsequent investigation of cluster-specific training methodologies that fundamentally restructure how chemical information is incorporated into the learning process.

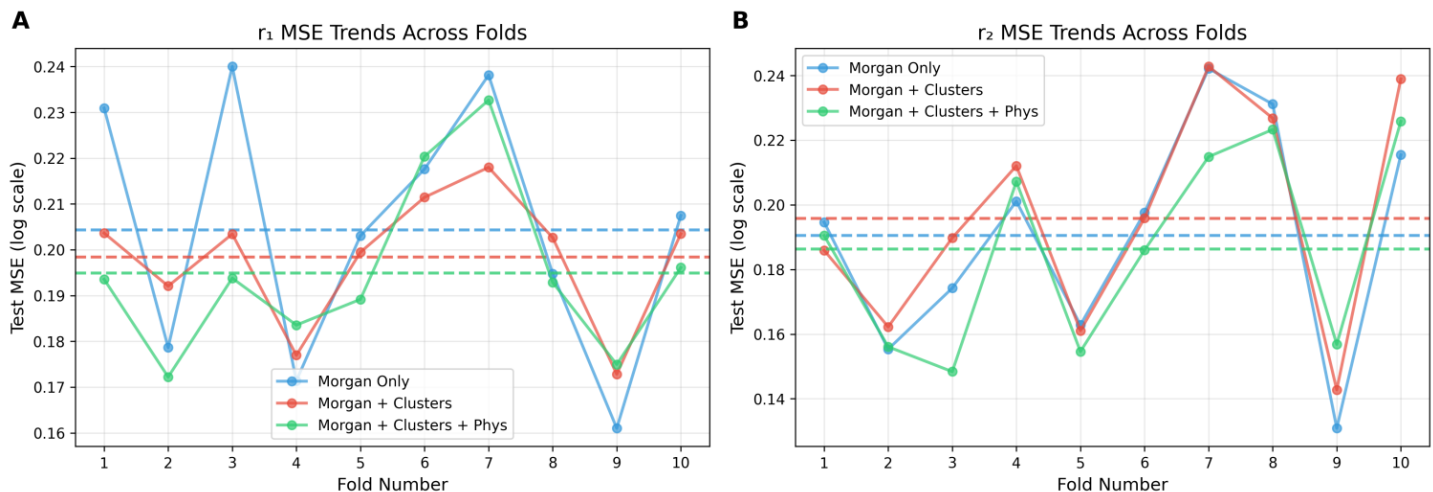

Figure 9. Test MSE trends across 10-fold cross-validation demonstrating incremental feature impact on reactivity ratio prediction. (A) $r_1$ MSE trends comparing three feature combinations: Morgan fingerprints only (blue), Morgan + cluster encoding (red), and Morgan + cluster + physicochemical descriptors (green). (B) $r_2$ MSE trends for the same three models. Dashed horizontal lines indicate the mean MSE across all folds for each model.

In contrast to the limited success of direct feature integration, the cluster-specific training approach demonstrated substantial improvements in predictive performance across both interaction types evaluated. As shown in Figure 10, for Cluster 1-1 interactions, comprising self-interactions within Cluster 1 monomers, the cluster-specific approach achieved test MSE values of $0.2525 \pm 0.0629$ for $r_1$ and $0.2626 \pm 0.0510$ for $r_2$ predictions in log scale, compared to $0.2845 \pm 0.0964$ and $0.2880 \pm 0.0545$ respectively for the general training approach. The $R^2$ score values also reflect this improvement, increasing from $0.246 \pm 0.160$ to $0.324 \pm 0.086$ for $r_1$ and from $0.214 \pm 0.121$ to $0.288 \pm 0.094$ for $r_2$. Notably, the cluster-specific model not only achieved higher $R^2$ values but also demonstrated significantly reduced variance across cross-validation folds, indicating more consistent and reliable predictions. The parity plots for individual cross-validation folds, presented in Supporting Information (SI), Section 6 (Figure S4 – Figure S14).

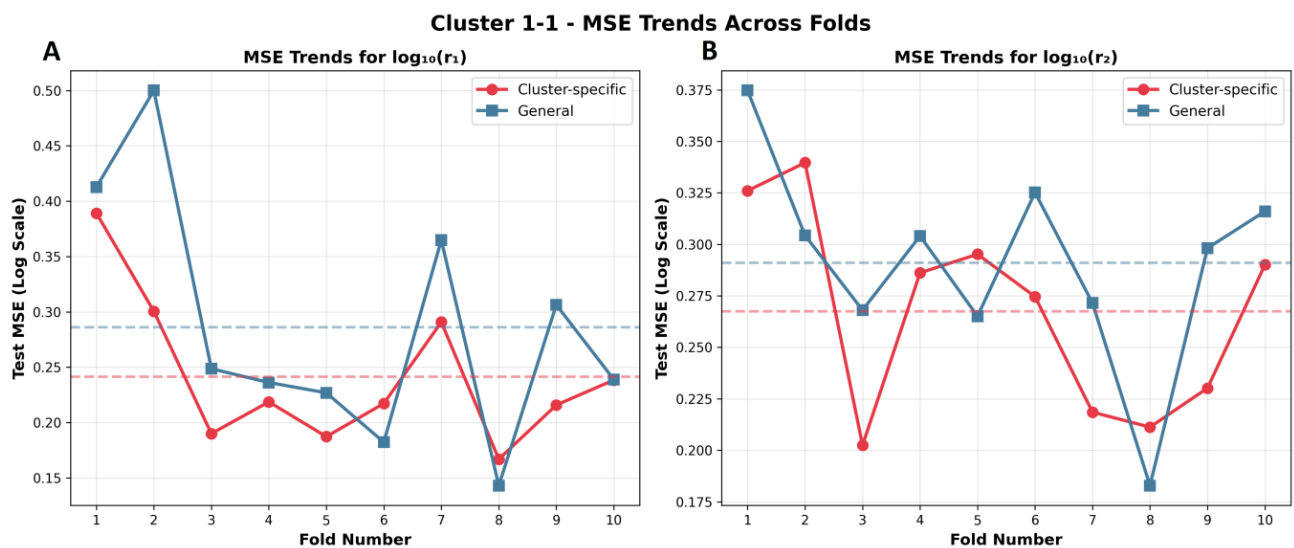

Figure 10. MSE trends across 10-fold cross-validation for Cluster 1-1 interactions comparing cluster-specific and general training approaches for (A) $r_1$ predictions, and (B) $r_2$ predictions. Cluster-specific training demonstrates lower MSE values and reduced variance across folds.

Expanding the evaluation to Cluster 1-2 interactions, which represent cross-cluster interactions between Cluster 1 and Cluster 2 monomers, the cluster-specific approach revealed even more substantial improvements in predictive performance. As shown in Figure 11, the cluster-specific models achieved test MSE values of $0.1976 \pm 0.0521$ for $r_1$ and $0.1947 \pm 0.0382$ for $r_2$ in log scale, compared to $0.2221 \pm 0.0562$ and $0.2333 \pm 0.0365$ respectively for the general training approach. The $R^2$ scores reflect this enhancement, increasing from $0.356 \pm 0.117$ to $0.429 \pm 0.103$ for $r_1$ and from $0.322 \pm 0.075$ to $0.429 \pm 0.120$ for $r_2$. Notably, the cluster-specific model demonstrated better performance with reduced variance across cross-validation folds for both reactivity ratios. These improvements demonstrate that cluster-specific training effectively captures the chemical relationships governing inter-cluster monomer interactions, with the fold-by-fold MSE trends and corresponding parity plots presented in Supporting Information (SI) Section 6 (Figure S15 – Figure S25).

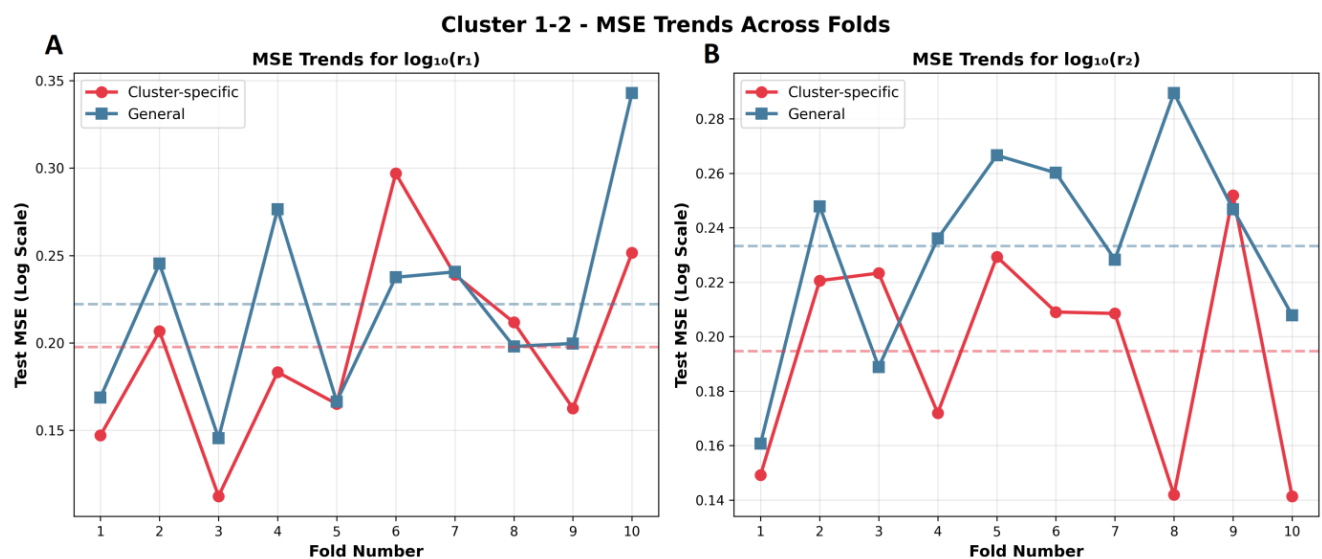

Figure 11. MSE trends across 10-fold cross-validation for Cluster 1-2 interactions comparing cluster-specific and general training approaches for (A) $r_1$ predictions, and (B) $r_2$ predictions. Cluster-specific training demonstrates lower MSE values and reduced variance across folds.

Data availability constraints prevented comprehensive evaluation across all six possible cluster interaction types, as only Cluster 1-1 (870 entries) and Cluster 1-2 (858 entries) interactions contained sufficient samples for robust neural network training and meaningful cross-validation analysis. The remaining cluster interactions had limited data that would not support reliable model development or validation. Despite this limitation, the consistent improvements observed across both tested interaction types—spanning both intra-cluster (1-1) and inter-cluster (1-2) interactions—suggest that cluster-specific training benefits would likely extend to other cluster combinations given sufficient experimental data.

These findings have important implications for experimental design strategies in polymer chemistry. Our results indicate that when developing predictive models for specific reactivity ratio predictions, targeted experiments within chemically homogeneous domains may yield superior model performance compared to broad experimental surveys across diverse monomer types. This suggests a shift in resource allocation strategy: rather than exhaustive screening across all possible monomer combinations, experimental efforts might be more effectively focused on systematic studies within specific cluster interactions relevant to the desired polymer architecture. The enhanced predictive accuracy achieved through cluster-specific training demonstrates that chemical focus and homogeneity in training data can improve model performance within the target domain, enabling more efficient experimental design and synthesis planning for specific copolymer applications.

## 4. CONCLUSION

This study presents a framework, integrating unsupervised learning-derived chemical information into supervised machine learning models for predicting reactivity ratio in chain radical copolymerization systems. Spectral clustering analysis of monomer physicochemical features led to the identification of three distinct chemical groups, revealing a characteristic reactivity patterns across six cluster interactions that correspond to diverse copolymer architectures—from highly predictable alternating sequences in Cluster 2-3 interactions ($r_1 \approx 0.5$, $r_2 \approx 0.1$ with minimal variance) to large configurational space in Cluster 1-1 interactions ($r_1$, $r_2 \approx 1.1$ with substantial variance) capable of producing configurations ranging from random to block copolymers. Investigation of two integrated methodologies revealed that while direct feature concatenation failed to improve performance beyond baseline models ($R^2 \approx 47\%$), cluster-specific training successfully leveraged chemical organization to achieve substantial improvements in both prediction accuracy and model stability for Cluster 1-1 and Cluster 1-2 interactions compared to general training approaches. The comparative analysis of unsupervised and supervised learning paradigms establishes their complementary roles in polymer informatics. Unsupervised learning (clustering) provides computationally efficient initial chemical organization and rapid trend identification across monomer families, serving as a valuable exploratory tool with minimal computational overhead. However, precise reactivity ratio predictions for specific monomer pairs necessitate supervised learning methods, particularly neural networks, which deliver better accuracy at the cost of increased computational demands. This dual-paradigm approach suggests an efficient hierarchical workflow: unsupervised clustering for strategic experimental planning and dataset organization, followed by targeted supervised learning for high-precision predictions in specific chemical domains.

The demonstrated advantages of cluster-specific training introduce important considerations for practical implementation that extend beyond purely computational metrics. While specialized models trained on chemically homogeneous subsets exhibited enhanced predictive accuracy and reduced cross-validation variance, these benefits are contingent upon sufficient data availability within each cluster interaction to ensure negligible partitioning effects. In data-rich scenarios, deploying multiple cluster-specific models can significantly outperform single general models by exploiting chemical homogeneity; conversely, in data-limited regimes, general training approaches may prove superior by maximizing information extraction from the complete dataset. These findings challenge conventional assumptions favoring comprehensive chemical diversity in training data, suggesting instead that targeted experimental efforts focused on specific cluster interactions—as identified through unsupervised analysis—may yield superior predictive performance for specific synthesis objectives. This paradigm shift has direct implications for experimental resource allocation in polymer chemistry: rather than pursuing exhaustive screening across all possible monomer combinations, strategic concentration of experimental efforts within relevant chemical domains, guided by unsupervised learning insights, can enhance both the efficiency of data collection and the accuracy of resulting predictive models. The integration of unsupervised chemical organization with supervised prediction thus establishes a practical framework for advancing machine learning applications in polymer synthesis, demonstrating that chemical knowledge, when appropriately encoded, can enhance model performance while informing optimal experimental design strategies.

## ASSOCIATED CONTENT

### Data Availability Statement

The complete collected dataset and all code implementations used in this study are publicly available at https://github.com/habibollah1994/reactivity-ratio-prediction

### Supporting Information

The supporting information is available free of charge. Section 1 provides comprehensive analysis of feature selection process for monomer representation, including rationale and criteria for molecular descriptor selection. Section 2 details the clustering optimization process, including evaluation metrics (silhouette score, Calinski-Harabasz index, and Davies-Bouldin index) for different clustering methods. Section 3 contains analysis of dimensionality reduction techniques and visualization approaches used for understanding monomer clustering patterns. Section 4 presents statistical data of interactions between monomer clusters. Section 5 presents comparative performance analysis of incremental feature addition to baseline models. Section 6 contains fold-by-fold parity plots for cluster-specific and general training models.

## AUTHOR INFORMATION

### Corresponding Authors

**Mona Bavarian** - Department of Chemical and Biomolecular Engineering, University of Nebraska-Lincoln, Lincoln, Nebraska, 68588, United States; https://orcid.org/0000-0001-7689-773X; Email: mona.bavarian@unl.edu

**Authors**

**Habibollah Safari** - Department of Chemical and Biomolecular Engineering, University of Nebraska−Lincoln, Lincoln, Nebraska 68588-8286, United States; https://orcid.org/0000-0003-1382-9582

**ACKNOWLEDGMENT**

The authors acknowledge the support of the National Science Foundation (NSF) under the Award Number 2238147. The authors would like to extend their thanks to the Holland Computing Center at the University of Nebraska-Lincoln for the essential computational resources and support provided, which played a crucial role in the conduct of this research.

**REFERENCES**

(1) Yeo, H.; Debnath, S.; Krishnan, B. P.; Boudouris, B. W. Radical Polymers in Optoelectronic and Spintronic Applications. *RSC Appl. Polym.* **2024**, *2* (1), 7–25. https://doi.org/10.1039/D3LP00213F.

(2) *Excipients in Pharmaceutical Additive Manufacturing: A Comprehensive Exploration of Polymeric Material Selection for Enhanced 3D Printing*. https://www.mdpi.com/1999-4923/16/3/317 (accessed 2024-10-21).

(3) Ismail, S.; Safari, H.; Bavarian, M. Design of Supported Ionic Liquid Membranes for CO2 Capture Using a Generative AI-Based Approach. *Ind. Eng. Chem. Res.* **2025**. https://doi.org/10.1021/acs.iecr.4c03280.

(4) Yang, G. G.; Choi, H. J.; Li, S.; Kim, J. H.; Kwon, K.; Jin, H. M.; Kim, B. H.; Kim, S. O. Intelligent Block Copolymer Self-Assembly towards IoT Hardware Components. *Nat. Rev. Electr. Eng.* **2024**, *1* (2), 124–138. https://doi.org/10.1038/s44287-024-00017-w.

(5) *Rational Design of New Conjugated Polymers with Main Chain Chirality for Efficient Optoelectronic Devices: Carbo[6]Helicene and Indacenodithiophene Copolymers as Model Compounds - Gedeon - 2024 - Advanced Materials - Wiley Online Library*. https://onlinelibrary.wiley.com/doi/full/10.1002/adma.202314337 (accessed 2024-10-21).

(6) Qiang, Y.; Xie, R.; He, B.; Liu, C.; Zhou, Q.; Wang, Y.; Gu, X.; Gong, X.; Liu, Y.; Liu, X. Designer Conjugation-Break Spacer That Boosts Charge Transport in Semiconducting Terpolymers. *Macromolecules* **2024**, *57* (12), 5902–5914. https://doi.org/10.1021/acs.macromol.4c00613.

(7) Zhu, Y.; Wu, H.; Martin, A.; Beck, P.; Allahyarov, E.; Wongwirat, T.; Rui, G.; Zhu, Y.; Hawthorne, D.; Fan, J.; Wu, J.; Zhang, S.; Zhu, L.; Kaur, S.; Pei, Q. Operando Investigation of the Molecular Origins of Dipole Switching in P(VDF-TrFE-CFE) Terpolymer for Large Adiabatic Temperature Change. *Adv. Funct. Mater.* **2024**, *34* (26), 2314705. https://doi.org/10.1002/adfm.202314705.

(8) Vagenas, D.; Pispas, S. Four-Component Statistical Copolymers by RAFT Polymerization. *Polymers* **2024**, *16* (10), 1321. https://doi.org/10.3390/polym16101321.

(9) Lundberg, D. J.; Kilgallon, L. J.; Cooper, J. C.; Starvaggi, F.; Xia, Y.; Johnson, J. A. Accurate Determination of Reactivity Ratios for Copolymerization Reactions with Reversible Propagation Mechanisms. *Macromolecules* **2024**, *57* (14), 6727–6740. https://doi.org/10.1021/acs.macromol.4c00835.

(10) McDonald, T.; Tsay, C.; Schweidtmann, A. M.; Yorke-Smith, N. Mixed-Integer Optimisation of Graph Neural Networks for Computer-Aided Molecular Design. arXiv December 2, 2023. https://doi.org/10.48550/arXiv.2312.01228.

(11) Mayo, F. R.; Walling, Cheves. Copolymerization. *Chem. Rev.* **1950**, *46* (2), 191–287. https://doi.org/10.1021/cr60144a001.

(12) *Demystifying the estimation of reactivity ratios for terpolymerization systems - Kazemi - 2014 - AIChE Journal - Wiley Online Library*. https://aiche.onlinelibrary.wiley.com/doi/full/10.1002/aic.14439 (accessed 2024-10-23).

(13) Scott, A. J.; Penlidis, A. Binary vs. Ternary Reactivity Ratios: Appropriate Estimation Procedures with Terpolymerization Data. *Eur. Polym. J.* **2018**, *105*, 442–450. https://doi.org/10.1016/j.eurpolymj.2018.06.021.

(14) Pujari, N. S.; Wang, M.; Gonsalves, K. E. Co and Terpolymer Reactivity Ratios of Chemically Amplified Resists. *Polymer* **2017**, *118*, 201–214. https://doi.org/10.1016/j.polymer.2017.05.001.

(15) *AMPS/AAm/AAc Terpolymerization: Experimental Verification of the EVM Framework for Ternary Reactivity Ratio Estimation*. https://www.mdpi.com/2227-9717/5/1/9 (accessed 2024-10-23).

(16) Huglin, M. B. The Reactivity Ratio in Styrene Maleic Anhydride Copolymerization. *Polymer* **1962**, *3*, 335–336. https://doi.org/10.1016/0032-3861(62)90089-7.

(17) Alfrey, T.; Lavin, E. The Copolymerization of Styrene and Maleic Anhydride. *J. Am. Chem. Soc.* **1945**, *67* (11), 2044–2045. https://doi.org/10.1021/ja01227a502.

(18) Iwakura, Y.; Kurosaki, T.; Nakabayashi, N. Reactive Fiber. Part I. Copolymerization and Copolymer of Acrylonitrile with Glycidyl Methacrylate and with Glycidyl Acrylate. *Makromol. Chem.* **1961**, *44* (1), 570–590. https://doi.org/10.1002/macp.1961.020440145.

(19) Kermagoret, A.; Mathieu, K.; Thomassin, J.-M.; Fustin, C.-A.; Duchêne, R.; Jérôme, C.; Detrembleur, C.; Debuigne, A. Double Thermoresponsive Di- and Triblock Copolymers Based on N-Vinylcaprolactam and N-Vinylpyrrolidone: Synthesis and Comparative Study of Solution Behaviour. *Polym. Chem.* **2014**, *5* (22), 6534–6544. https://doi.org/10.1039/C4PY00852A.

(20) You, H.; Tirrell, D. A. Radical Copolymerization of 2-Ethylacrylic Acid and Methacrylic Acid. *J. Polym. Sci. Part Polym. Chem.* **1990**, *28* (11), 3155–3163. https://doi.org/10.1002/pola.1990.080281121.

(21) Šorm, M.; Nešpůrek, S. Alkali Salts of (ω-Sulphoxyalkyl)-Acrylates and -Methacrylates: Radical Polymerization and Copolymerization. *Eur. Polym. J.* **1978**, *14* (12), 977–980. https://doi.org/10.1016/0014-3057(78)90153-2.

(22) Webb, M. A.; Jackson, N. E.; Gil, P. S.; de Pablo, J. J. Targeted Sequence Design within the Coarse-Grained Polymer Genome. *Sci. Adv.* **2020**, *6* (43), eabc6216. https://doi.org/10.1126/sciadv.abc6216.

(23) Zhou, H.; Fang, Y.; Li, L.; Liu, P.; Gao, H. De Novo Design of Polymers with Specified Properties Using Reinforcement Learning. *Macromolecules* **2025**, *58* (11), 5477–5486. https://doi.org/10.1021/acs.macromol.5c00427.

(24) Zheng, Y.; Biswal, A. K.; Guo, Y.; Thakolkaran, P.; Kokane, Y.; Varshney, V.; Kumar, S.; Vashisth, A. Toward Sustainable Polymer Design: A Molecular Dynamics-Informed Machine Learning Approach for Vitrimers. *Digit. Discov.* **2025**. https://doi.org/10.1039/D5DD00239G.

(25) Zhao, W.; Xu, X.; Lan, H.; Wang, L.; Lin, J.; Du, L.; Zhang, C.; Tian, X. Designing Multicomponent Thermosetting Resins through Machine Learning and High-Throughput Screening. *Macromolecules* **2025**, *58* (1), 744–753. https://doi.org/10.1021/acs.macromol.4c01822.

(26) Nguyen, T.; Bavarian, M. Machine Learning Approach to Polymer Reaction Engineering: Determining Monomers Reactivity Ratios. *Polymer* **2023**, *275*, 125866. https://doi.org/10.1016/j.polymer.2023.125866.

(27) Dossi, M.; Moscatelli, D. A QM Approach to the Calculation of Reactivity Ratios in Free-Radical Copolymerization. *Macromol. React. Eng.* **2012**, *6* (2–3), 74–84. https://doi.org/10.1002/mren.201100065.

(28) Farajzadehahary, K.; Telleria-Allika, X.; Asua, J. M.; Ballard, N. An Artificial Neural Network to Predict Reactivity Ratios in Radical Copolymerization. *Polym. Chem.* **2023**, *14* (23), 2779–2787. https://doi.org/10.1039/D3PY00246B.

(29) Safari, H.; Bavarian, M. Enhancing Polymer Reaction Engineering Through the Power of Machine Learning. *Syst. Control Trans.* **2024**, 157792.

(30) Takahashi, K.; Mamitsuka, H.; Tosaka, M.; Zhu, N.; Yamago, S. CoPolDB: A Copolymerization Database for Radical Polymerization. *Polym. Chem.* **2024**, *15* (10), 965–971. https://doi.org/10.1039/D3PY01372C.

(31) Jackson, N. E.; Savoie, B. M. Ten Problems in Polymer Reactivity Prediction. *Macromolecules* **2025**, *58* (4), 1737–1754. https://doi.org/10.1021/acs.macromol.4c02582.

(32) Lewis, F. M.; Walling, C.; Cummings, W.; Briggs, E. R.; Mayo, F. R. Copolymerization. IV. Effects of Temperature and Solvents on Monomer Reactivity Ratios. *J. Am. Chem. Soc.* **1948**, *70* (4), 1519–1523. https://doi.org/10.1021/ja01184a066.

(33) Fernández-García, M.; Fernández-Sanz, M.; Madruga, E. L.; Cuervo-Rodriguez, R.; Hernández-Gordo, V.; Fernández-Monreal, M. C. Solvent Effects on the Free-Radical Copolymerization of Styrene with Butyl Acrylate. I. Monomer Reactivity Ratios. *J. Polym. Sci. Part Polym. Chem.* **2000**, *38* (1), 60–67. https://doi.org/10.1002/(SICI)1099-0518(20000101)38:1<60::AID-POLA8>3.0.CO;2-F.

(34) Yamada, K.; Nakano, T.; Okamoto, Y. Free-Radical Copolymerization of Vinyl Esters Using Fluoroalcohols as Solvents: The Solvent Effect on the Monomer Reactivity Ratio. *J. Polym. Sci. Part Polym. Chem.* **2000**, *38* (1), 220–228. https://doi.org/10.1002/(SICI)1099-0518(20000101)38:1<220::AID-POLA27>3.0.CO;2-P.

(35) Hou, C.; Liu, J.; Ji, C.; Ying, L.; Sun, H.; Wang, C. Monomer Apparent Reactivity Ratios for Acrylonitrile/Methyl Vinyl Ketone Copolymerization System. *J. Appl. Polym. Sci.* **2006**, *102* (4), 4045–4048. https://doi.org/10.1002/app.24328.

(36) He, Y.; Luscombe, C. K. Quantitative Comparison of the Copolymerisation Kinetics in Catalyst-Transfer Copolymerisation to Synthesise Polythiophenes. *Polym. Chem.* 15 (25), 2598–2605. https://doi.org/10.1039/d4py00009a.

(37) Fischer, E. J.; Cuccato, D.; Storti, G.; Morbidelli, M. Effect of the Charge Interactions on the Composition Behavior of Acrylamide/Acrylic Acid Copolymerization in Aqueous Medium. *Eur. Polym. J.* **2018**, *98*, 302–312. https://doi.org/10.1016/j.eurpolymj.2017.11.022.

(38) Morgenthaler, E. C.; Ribbe, A. E.; Bradley, L. C.; Emrick, T. Alkyne-Rich Patchy Polymer Colloids Prepared by Surfactant-Free Emulsion Polymerization. *J. Colloid Interface Sci.* **2025**, *679*, 276–283. https://doi.org/10.1016/j.jcis.2024.10.040.

(39) Wiley, R. H.; Sale, E. E. Tracer Techniques for the Determination of Monomer Reactivity Ratios. II. Monomer Reactivity Ratios in Copolymerizations with Divinyl Monomers. *J. Polym. Sci.* **1960**, *42* (140), 491–500. https://doi.org/10.1002/pol.1960.1204214016.

(40) Storey, B. T. Copolymerization of Styrene and P-Divinylbenzene. Initial Rates and Gel Points. *J. Polym. Sci. A* **1965**, *3* (1), 265–282. https://doi.org/10.1002/pol.1965.100030128.

(41) Usefi, H. Clustering, Multicollinearity, and Singular Vectors. *Comput. Stat. Data Anal.* **2022**, *173*, 107523. https://doi.org/10.1016/j.csda.2022.107523.

(42) MacQueen, J. Some Methods for Classification and Analysis of Multivariate Observations. In *Proceedings of the Fifth Berkeley Symposium on Mathematical Statistics and Probability, Volume 1: Statistics*; University of California Press, 1967; Vol. 5.1, pp 281–298.

(43) Ng, A.; Jordan, M.; Weiss, Y. On Spectral Clustering: Analysis and an Algorithm. *Adv. Neural Inf. Process. Syst.* **2001**, *14*.

(44) Ward Jr., J. H. Hierarchical Grouping to Optimize an Objective Function. *J. Am. Stat. Assoc.* **1963**, *58* (301), 236–244. https://doi.org/10.1080/01621459.1963.10500845.

(45) Dempster, A. P.; Laird, N. M.; Rubin, D. B. Maximum Likelihood from Incomplete Data via the EM Algorithm. *J. R. Stat. Soc. Ser. B Methodol.* **1977**, *39* (1), 1–38.

(46) Zhang, T.; Ramakrishnan, R.; Livny, M. BIRCH: An Efficient Data Clustering Method for Very Large Databases. *SIGMOD Rec* **1996**, *25* (2), 103–114. https://doi.org/10.1145/235968.233324.

(47) Rousseeuw, P. J. Silhouettes: A Graphical Aid to the Interpretation and Validation of Cluster Analysis. *J. Comput. Appl. Math.* **1987**, *20*, 53–65. https://doi.org/10.1016/0377-0427(87)90125-7.

(48) Calinski, T.; Harabasz, J. A Dendrite Method for Cluster Analysis. *Commun. Stat. - Theory Methods* **1974**, *3* (1), 1–27. https://doi.org/10.1080/03610927408827101.

(49) Davies, D. L.; Bouldin, D. W. A Cluster Separation Measure. *IEEE Trans. Pattern Anal. Mach. Intell.* **1979**, *PAMI-1* (2), 224–227. https://doi.org/10.1109/TPAMI.1979.4766909.

(50) Jolliffe, I. T.; Cadima, J. Principal Component Analysis: A Review and Recent Developments. *Philos. Trans. R. Soc. Math. Phys. Eng. Sci.* **2016**, *374* (2065), 20150202. https://doi.org/10.1098/rsta.2015.0202.

(51) Maaten, L. van der; Hinton, G. Visualizing Data Using T-SNE. *J. Mach. Learn. Res.* **2008**, *9* (86), 2579–2605.

(52) McInnes, L.; Healy, J.; Melville, J. UMAP: Uniform Manifold Approximation and Projection for Dimension Reduction. arXiv September 18, 2020. https://doi.org/10.48550/arXiv.1802.03426.

(53) Morgan, H. L. The Generation of a Unique Machine Description for Chemical Structures-A Technique Developed at Chemical Abstracts Service. *J. Chem. Doc.* **1965**, *5* (2), 107–113. https://doi.org/10.1021/c160017a018.

(54) Rogers, D.; Hahn, M. Extended-Connectivity Fingerprints. *J. Chem. Inf. Model.* **2010**, *50* (5), 742–754. https://doi.org/10.1021/ci100050t.

# Supporting Information

# Chemically-Informed Machine Learning Approach for Prediction of Reactivity Ratios in Radical Copolymerization

Habibollah Safari, Mona Bavarian

Department of Chemical and Biomolecular Engineering, University of Nebraska-Lincoln,
Lincoln, Nebraska 68588, United States

Affiliations: Corresponding Author: mona.bavarian@unl.edu

**Section 1. Features and Monomer Descriptor Selection Criteria**

A comprehensive analysis was performed for the selection of appropriate descriptors and features of monomers through which all important aspects of physicochemical properties of monomers in radical polymerization are represented. The initial selection process considered fifteen molecular descriptors, each potentially significant in determining monomer reactivity and polymerization behavior. These descriptors were systematically evaluated through both statistical correlation analysis and chemical reasoning to ensure comprehensive coverage of relevant molecular properties while minimizing redundancy in the final feature set. The details of evaluation are presented in the following.

1. The analysis began with molecular weight, a fundamental size parameter showing significant correlations with several other descriptors. The correlation matrix revealed strong relationships with molecular volume (0.91) and number of connected hydrogens (0.71), as well as moderate correlations with total polar surface area (0.48) and molecular LogP (0.57). Despite these correlations, molecular weight was retained in the final feature set due to its fundamental importance in polymerization kinetics, particularly in diffusion-controlled processes. Furthermore, molecular weight serves as a critical indicator of molecular size, directly influencing spatial arrangements and steric effects during propagation. Moreover, molecular weight offers superior experimental accessibility and measurement reliability compared to related parameters, making it an essential descriptor for practical applications in polymer synthesis. Molecular volume, while providing important information about spatial effects in polymerization, showed strong correlations with multiple parameters. Beyond its high correlation with molecular weight (0.91), it also strongly correlated with number of connected hydrogens (0.86) and showed moderate correlation with hybridization sp³ (0.68). These multiple strong correlations, combined with its computational method dependency and the fact that its key information is largely captured by molecular weight, led to its elimination from the final feature set.

2. Since radical polymerization fundamentally occurs at the vinyl group position, inclusion of information about this functional group is critical for understanding and predicting reaction behavior. The vinyl group characteristics were analyzed through two key parameters: vinyl position (in loop versus linear) and vinyl carbons charge. The vinyl position parameter demonstrated remarkable independence, with correlation coefficients not exceeding 0.25 for any other descriptor. This statistical independence, combined with its crucial role in determining monomer reactivity (as evidenced by the significant difference in reactivity ratios between ringed and unringed configurations), made it an essential inclusion in the final feature set. Similarly, vinyl carbons charge showed minimal correlations with other parameters (maximum correlation of 0.25 with vinyl position) while providing critical information about electronic effects at the reaction center, justifying its retention.
3. Hybridization states were represented by three parameters: sp, sp², and sp³ hybridization. The sp² hybridization showed a very strong correlation with conjugated bonds (0.95) but maintained relatively low correlations with other parameters. Given its fundamental importance in vinyl polymerization and its more direct relationship with reactivity compared to conjugated bonds, sp² hybridization was retained. In contrast, sp³ hybridization showed high correlations with number of connected hydrogens (0.91) and moderate correlation with molecular volume (0.68), leading to its elimination. The sp hybridization, while showing low correlations across the board, was eliminated due to its limited relevance in vinyl polymerization systems, where sp² centers dominate reactivity.
4. Surface polarity and interaction capabilities were assessed through several parameters. Total polar surface area (TPSA) exhibited a strong correlation with number of hydrogen acceptors (0.86) but was retained as it provides a more comprehensive measure of surface polarity and better represents overall interaction capabilities. The number of hydrogen acceptors, despite its importance in hydrogen bonding, was eliminated due to this high correlation with TPSA and its more limited scope of information.
5. The number of hydrogen donors emerged as a remarkably independent parameter, showing low correlations across all other descriptors (maximum correlation of 0.44 with TPSA). This independence, combined with its crucial role in chain transfer reactions and hydrogen bonding effects during polymerization, secured its place in the final feature set.
6. Molecular LogP, representing the hydrophobic/hydrophilic balance of monomers, showed moderate correlations with several parameters (maximum 0.62 with molecular volume) but was retained due to its unique contribution to understanding phase behavior and monomer-monomer interactions in polymerization systems. This parameter provides essential information about monomer distribution in heterogeneous systems that is not captured by other descriptors.

7. The number of connected hydrogens showed significant correlations with molecular weight (0.71), molecular volume (0.86), and $sp^3$ hybridization (0.91). These multiple strong correlations, combined with the fact that its effects are largely captured by other selected parameters, led to its elimination from the final set.
8. Conjugated bonds, showing very high correlation with $sp^2$ hybridization (0.95), were eliminated as this electronic effect is better represented by the more fundamental $sp^2$ hybridization parameter. While conjugation plays a crucial role in radical stability and propagation kinetics, its effects are effectively captured through the combination of $sp^2$ hybridization and vinyl carbon charge parameters.
9. Finally, stereochemistry and chirality showed perfect correlation (1.00) with each other, necessitating the selection of one parameter. Stereochemistry was retained over chirality as it provides a more comprehensive description of spatial arrangements and their influence on propagation.

Through this systematic evaluation, eight descriptors were selected for the final feature set: molecular weight, vinyl position, vinyl carbons charge, $sp^2$ hybridization, total polar surface area, molecular LogP, number of hydrogen donors, and stereochemistry. This carefully curated set minimizes redundancy while maintaining comprehensive coverage of the physicochemical properties governing radical polymerization behavior. The selected features provide independent yet complementary information about monomer characteristics, creating a robust foundation for subsequent clustering analysis and reactivity ratio prediction.

      To validate our feature selection process, we employed Principal Component Analysis (PCA) in two complementary stages. First, we analyzed the complete set of fifteen original molecular descriptors to understand the inherent redundancy in our initial feature set. Principal Component Analysis transforms our original features into new uncorrelated variables (principal components) that capture the main patterns of variation in our data. In this context, variance represents how much our molecular descriptors vary across different monomers in the dataset, with cumulative variance indicating how much of this total variation we can capture as we add more principal components. Figure S1a shows the cumulative explained variance ratio for all fifteen original features. The analysis reveals that eight principal components are sufficient to explain 95% of the total variance in the original dataset (indicated by the horizontal dashed line). This finding suggests significant redundancy in our initial set of fifteen features, as we can capture most of the meaningful variation with approximately half the number of components. However, it's crucial to understand that these eight principal components are mathematical combinations of all fifteen features and do not directly correspond to our selected molecular descriptors. Rather, this analysis provided initial evidence that a carefully chosen set of eight features might be sufficient to describe the essential characteristics of our monomers. To validate our specific feature selection, we performed a second PCA analysis on our eight selected features (Figure S1b): molecular

weight, vinyl position, vinyl carbons charge, sp² hybridization, total polar surface area, molecular LogP, number of hydrogen donors, and stereochemistry. These features were chosen based on both correlation analysis and chemical understanding, as detailed in the previous sections. The cumulative variance plot for these eight features reveals that we need seven components to reach 95% of the total variance, indicating that our selected features are indeed capturing distinct aspects of monomer properties.

Figure S1. PCA validation of molecular descriptor selection. (a) Cumulative explained variance ratio for the original fifteen molecular descriptors, showing that eight principal components capture 95% of the total variance, indicating redundancy in the initial feature set. (b) Cumulative explained variance ratio for the eight selected features, demonstrating that seven components are needed to reach 95% variance, validating the independence of our chosen molecular descriptors. The dashed line represents the 95% variance threshold.

The fact that we need seven out of eight possible components to explain 95% of the variance provides strong validation of our feature selection strategy. It confirms that each selected feature contributes unique and important information about monomer properties, with minimal redundancy in the final set. This mathematical validation complements our chemical reasoning and correlation analysis, demonstrating that we have successfully identified a set of features that efficiently capture different aspects of monomer characteristics relevant to polymerization behavior.

**Section 2. Clustering Optimization**

To determine the optimal clustering approach for monomer clustering, we conducted a comprehensive evaluation of various clustering algorithms using three distinct evaluation metrics. Figure S2 presents the comparative performance of k-means, Spectral Clustering, Agglomerative Clustering, Gaussian Mixture Model (GMM), and BIRCH algorithms across Silhouette score, Calinski-Harabasz index, and Davies-Bouldin index.

The Silhouette score, illustrated in Figure S2a, was employed to evaluate the quality of our clustering. Formally, the Silhouette coefficient s(i) for each monomer i is given by:

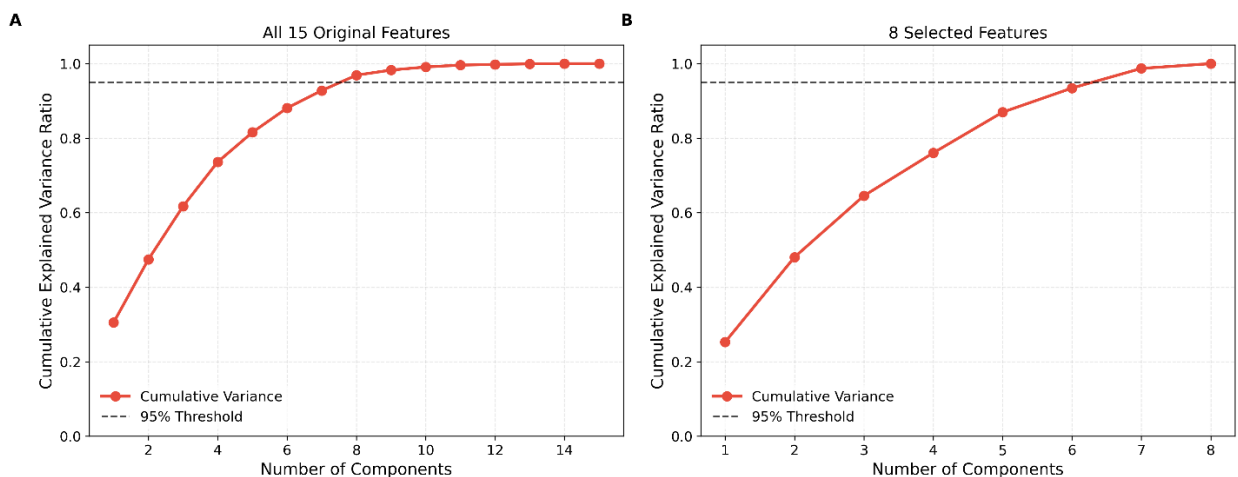

$$s(i) = \frac{b(i) - a(i)}{\max\{a(i), b(i)\}}$$

where a(i) is the average distance between monomer i and all other monomers in the same cluster, and b(i) is the minimum average distance from monomer i to any other cluster. A high (positive) s(i) indicates that monomer i is well matched to its own cluster and poorly matched to neighboring clusters. To obtain an overall Silhouette score for a clustering solution, we compute the mean of s(i) across all monomers in the dataset. This single value, which ranges from −1 to +1, summarizes the overall cohesion (similarity of monomers within each cluster) and separation (difference from other clusters). As evident in Figure S2a, Spectral clustering (red line) achieves consistently higher Silhouette scores (approximately 0.95-1.0) across different numbers of clusters, highlighting its notable stability. By contrast, other algorithms—such as BIRCH (purple line) and GMM (yellow line)—demonstrate significant performance declines as the number of clusters increases, especially beyond k=5. This difference underscores the strong intra-cluster cohesion and inter-cluster separation delivered by Spectral clustering with our chosen feature set.

The Calinski-Harabasz (CH) index, presented in Figure S2b, evaluates clustering quality by comparing within-cluster dispersion to between-cluster dispersion. Formally, for k clusters and n total points, the CH index can be expressed as:

$$CH = \frac{SS_B/(k-1)}{SS_W/(n-k)}$$

where $SS_B$ is the sum of squares between clusters, and $SS_W$ is the sum of squares within clusters. Higher CH values generally indicate better defined, more distinct clusters. In this study, while Agglomerative clustering (blue line) shows increasing performance with higher k—maintaining scores above 0.90 beyond k=3—Spectral clustering exhibits a characteristic pattern: strong initial performance followed by stabilization around 0.55. This pattern suggests that Spectral clustering efficiently captures the inherent structure of the molecular data with moderate cluster separation, even though its CH scores are lower than some other methods.

The Davies-Bouldin (DB) index, illustrated in Figure S2c, measures the average similarity between each cluster and its most similar neighbor; lower values denote better separation. Mathematically, for k clusters, the DB index is given by:

$$DB = \frac{1}{k}\sum_{i=1}^{k} \max_{j \neq i}\left(\frac{s_i + s_j}{d(c_i, c_j)}\right)$$

where $s_i$ and $s_j$ are the average within-cluster distances (i.e., the "scatter") for clusters i and j, and $d(c_i, c_j)$ is the distance between the respective cluster centers. Our results show Spectral clustering maintaining moderate, stable DB values (approximately 0.65), indicating fairly consistent separation. By contrast, other methods—such as K-means (green line) and Agglomerative clustering—display more erratic trends, with scores rising from near 0 to above 0.4 for larger k, signaling reduced cluster distinction as the number of clusters grows.

Based on our comprehensive analysis of clustering metrics shown in Figure S2, Spectral clustering with three clusters emerges as the optimal choice for monomer classification. This determination is supported by multiple performance indicators: the algorithm maintains exceptionally high and stable Silhouette scores (approximately 0.95-1.0), demonstrates consistent Davies-Bouldin indices (around 0.65), and shows steady Calinski-Harabasz performance. While other algorithms like K-means and Agglomerative clustering show competitive performance in individual metrics, they lack the consistent performance across all three evaluation criteria that Spectral clustering achieves. The choice of three clusters is particularly compelling as it represents an optimal balance point – beyond this number, most algorithms show diminishing returns or performance degradation, as evidenced by the plateauing or declining metric scores.

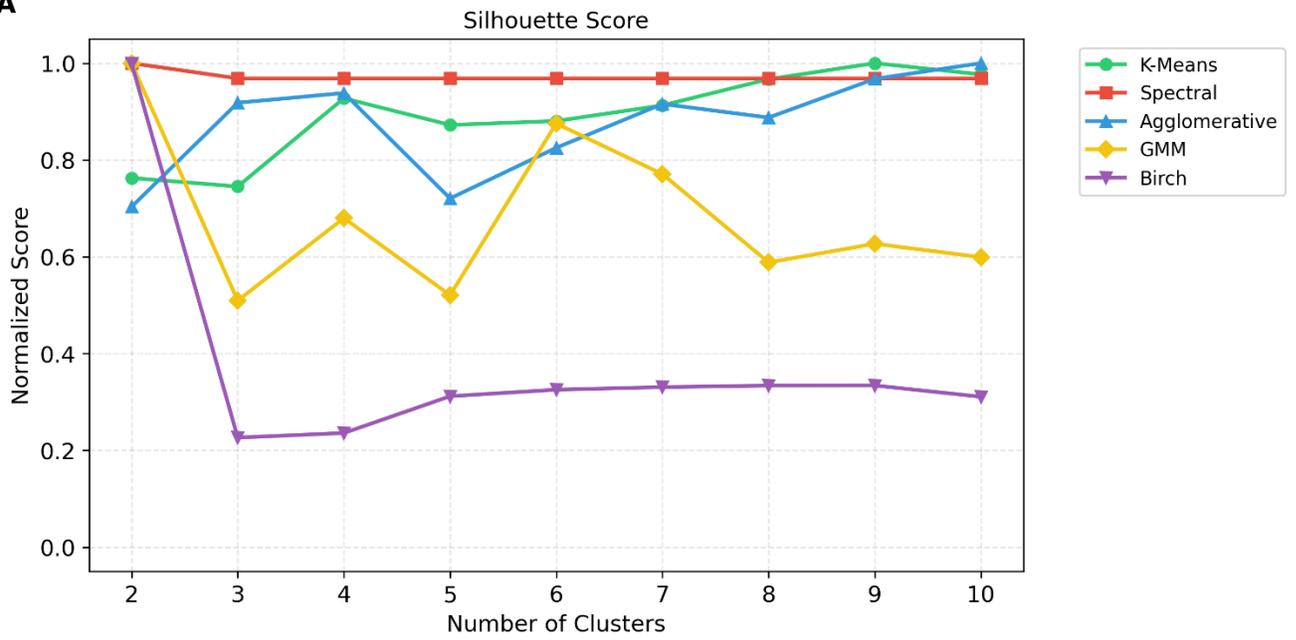

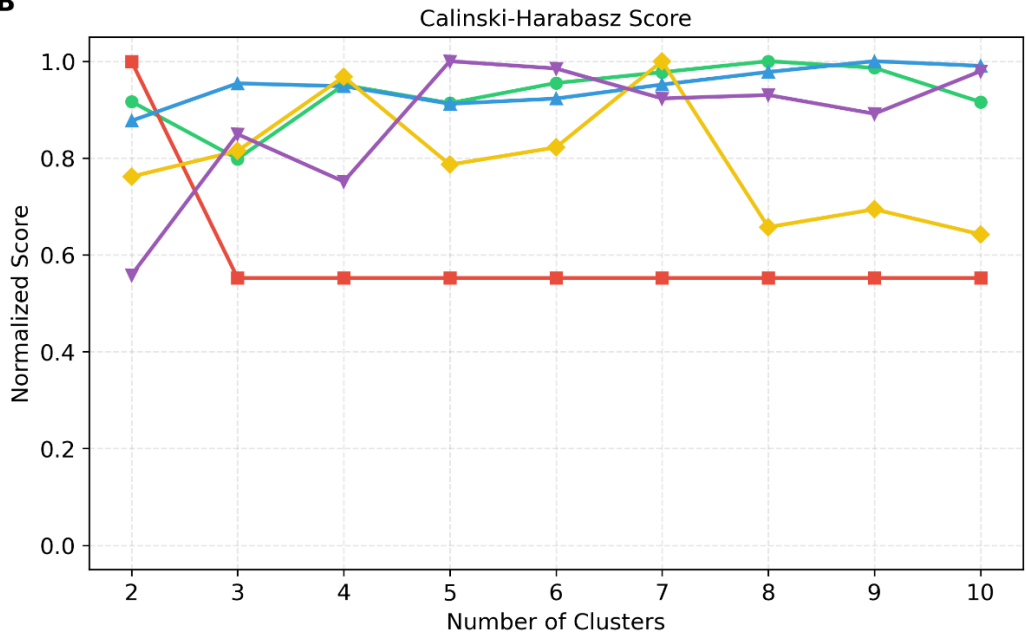

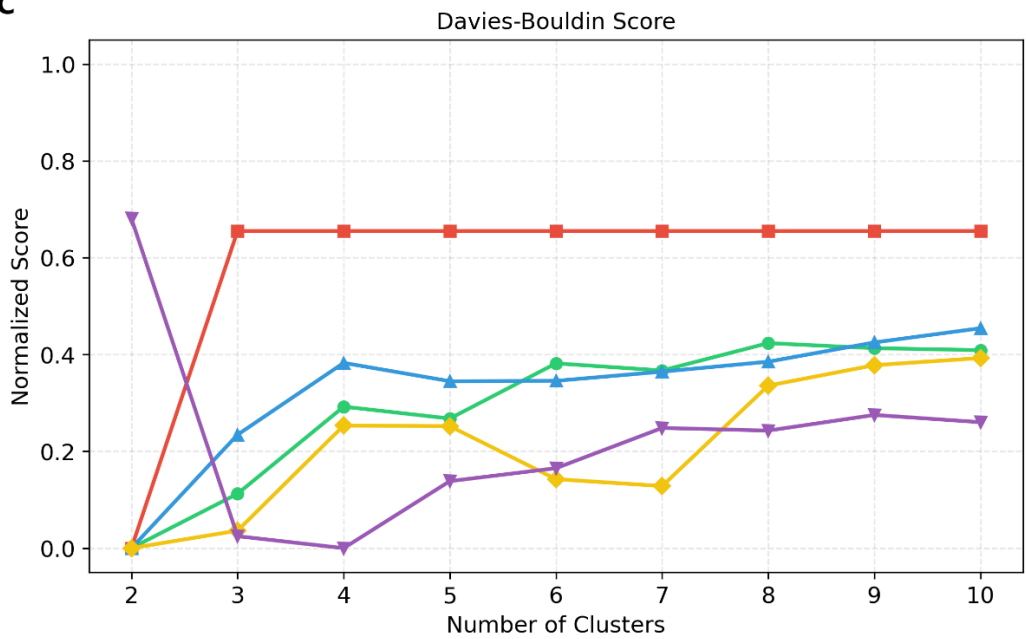

Figure S2. Comparison of clustering algorithms performance evaluated through different metrics. (a) Silhouette score (higher is better) measures cluster cohesion and separation, (b) Calinski-Harabasz index (higher is better) assesses the ratio of between-cluster to within-cluster variance, and (c) Davies-Bouldin index (lower is better) evaluates average similarity between clusters. All scores are normalized for comparative visualization. Algorithms were evaluated across cluster numbers ranging from 2 to 10.

## Section 3. Dimensionality Reduction and Visualization

To visualize and understand the high-dimensional molecular feature space, we employed three complementary dimensionality reduction techniques: Principal Component Analysis (PCA), t-distributed Stochastic Neighbor Embedding (t-SNE), and Uniform Manifold Approximation and Projection (UMAP). Figure S3 presents both the raw data distribution and clustered visualizations in two dimensions. The three methods reveal markedly different visualization characteristics that highlight distinct aspects of the data structure. PCA (Panels A and D) demonstrates significant cluster overlap. In contrast, t-SNE (Panels B and E) exhibits the most pronounced cluster separation, with Cluster 1 forming the largest distributed group across multiple regions, Cluster 2 appearing as distinct, well-separated subgroups in various portions of the embedding space, and Cluster 3 manifesting as a compact, tightly-bound group. UMAP (Panels C and F) provides an intermediate visualization approach, achieving clearer separation than PCA while maintaining more structured global organization than t-SNE, as evidenced by the organized cluster distribution and preserved neighborhood relationships.

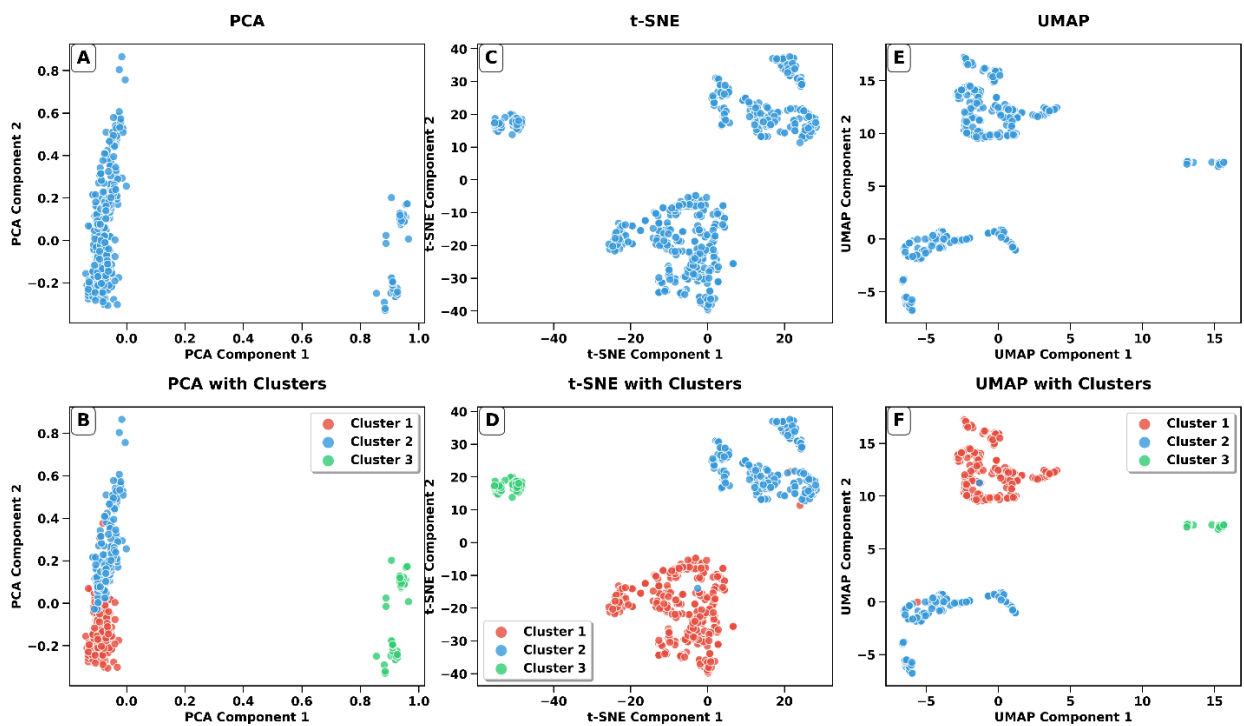

Figure S3. Dimensionality reduction and clustering visualization of monomer data. Comparison of PCA (A, B), t-SNE (C,D), and UMAP (E, F) techniques. The consistent cluster separation across all methods, particularly in t-SNE projections, demonstrates the robustness of the identified monomer groups.

## Section 4. Statistical Analysis of Interaction of Between Clusters

The interaction between different monomer clusters reveals distinct patterns that can be quantitatively analyzed through their reactivity ratio statistics, as shown in Table S1. Based on these data, each cluster demonstrates unique polymerization behavior that directly influences the resulting polymer chain configuration. Analysis of cluster interactions through reactivity ratio patterns ($\mu_r$, $\sigma_r$) provides valuable insights into the likely arrangement of monomers in the resulting polymer chains.

Table S1. Statistical analysis of reactivity ratio patterns across monomer cluster interactions. Higher mean values suggest block-like sequences, values near 1.0 suggest random incorporation, and lower values suggest alternating sequences.

| ($\mu_r$, $\sigma_r$) | Cluster 1 | Cluster 2 | Cluster 3 |
|---|---|---|---|
| Cluster 1 | C1: (1.097, 1.482) | C1:(0.631, 0.968) C2:(1.134, 1.623) | C1:(1.375, 1.770) C3:(0.707, 1.340) |
| Cluster 2 | C1:(0.631, 0.968) C2:(1.134, 1.623) | C2:(0.817, 1.014) | C2:(0.526, 0.887) C3:(0.092, 0.094) |
| Cluster 3 | C1:(1.375, 1.770) C3:(0.707, 1.340) | C2:(0.526, 0.887) C3:(0.092, 0.094) | C3:(0.603, 0.546) |

Cluster 1 self-interactions exhibit mean reactivity ratios slightly above unity ($\mu_r = 1.097$) with the high standard deviation ($\sigma_r = 1.482$), indicating a tendency toward random-to-block copolymer formation with considerable variability in behavior depending on the specific monomer pair. This high variance reflects the diverse chemical nature of monomers within Cluster 1, encompassing a broad range of reactivity characteristics. Cross-cluster interactions reveal pronounced asymmetric patterns that suggest gradient copolymer formation. Cluster 1-Cluster 2 interactions show moderate asymmetry, with Cluster 1 exhibiting lower reactivity ($\mu_r = 0.631$) while Cluster 2 demonstrates higher values ($\mu_r = 1.134$). More dramatically, Cluster 1-Cluster 3 interactions display the highest mean reactivity ratio observed ($\mu_r = 1.375$ for Cluster 1) paired with moderate values for Cluster 3 ($\mu_r = 0.707$), indicating strong potential for gradient sequence development. The most distinctive pattern emerges in Cluster 2-Cluster 3 interactions, where Cluster 3 exhibits extremely low reactivity ratios ($\mu_r = 0.092$) with minimal standard deviation ($\sigma_r = 0.094$), strongly indicating alternating copolymer formation with highly predictable sequence arrangements. This remarkably low variance suggests consistent alternating behavior across different monomer pairs within this interaction type. Cluster 2 also shows reduced reactivity in this pairing ($\mu_r = 0.526$), further supporting the alternating tendency. Cluster 3 self-interactions demonstrate moderate reactivity ratios ($\mu_r = 0.603$) with the lowest standard deviation among all self-interactions ($\sigma_r = 0.546$), suggesting consistent random copolymer formation. These distinct cluster behaviors and their quantified interactions provide a framework for predicting chain configurations and guiding monomer selection in targeted copolymer synthesis, while the substantial standard deviations observed in most interaction types underscore the importance of individual monomer pair predictions rather than relying solely on cluster averages.

**Section 5. Comparative Performance Analysis: Incremental Addition of Cluster and Physicochemical Features to Morgan Fingerprints**

Table S2 presents comprehensive performance metrics for three feature integration models evaluated through 10-fold cross-validation. The Morgan fingerprints baseline established the initial performance benchmark. Adding cluster encoding resulted in minimal improvement over the baseline model. The complete feature set incorporating both cluster and physicochemical descriptors showed modest gains in predictive performance.

Table S2. Comprehensive performance metrics for three feature integration models across 10-fold cross-validation.

| Model | Train Mean Square Error ($Log_{10} r_1$) | Train Mean Square Error ($Log_{10} r_2$) | Test Mean Square Error ($Log_{10} r_1$) | Test Mean Square Error ($Log_{10} r_2$) | Train $R^2$ Score ($Log_{10} r_1$) | Train $R^2$ Score ($Log_{10} r_2$) | Test $R^2$ Score ($Log_{10} r_1$) | Test $R^2$ Score ($Log_{10} r_2$) | Test Mean Square Error ($r_1$) | Test Mean Square Error ($r_2$) | Test $R^2$ Score ($r_1$) | Test $R^2$ Score ($r_2$) |
|---|---|---|---|---|---|---|---|---|---|---|---|---|
| Morgan Only | 0.0339 ± 0.0099 | 0.0331 ± 0.0115 | 0.2043 ± 0.0279 | 0.1905 ± 0.0349 | 0.9080 ± 0.0271 | 0.9100 ± 0.0317 | 0.4435 ± 0.0610 | 0.4782 ± 0.0751 | 1.5061 ± 0.4753 | 1.3941 ± 0.4064 | 0.2000 ± 0.1063 | 0.2500 ± 0.0525 |
| Morgan + Cluster | 0.0327 ± 0.0097 | 0.0299 ± 0.0060 | 0.1984 ± 0.0142 | 0.1958 ± 0.0343 | 0.9112 ± 0.0261 | 0.9189 ± 0.0162 | 0.4572 ± 0.0486 | 0.4641 ± 0.0685 | 1.4471 ± 0.3746 | 1.4165 ± 0.4328 | 0.2290 ± 0.0588 | 0.2385 ± 0.0661 |
| Morgan + Cluster + Physiochemical Features | 0.0359 ± 0.0137 | 0.0328 ± 0.0118 | 0.1949 ± 0.0187 | 0.1863 ± 0.0306 | 0.9028 ± 0.0361 | 0.9108 ± 0.0320 | 0.4669 ± 0.0557 | 0.4900 ± 0.0572 | 1.4208 ± 0.4476 | 1.3970 ± 0.4076 | 0.2539 ± 0.0803 | 0.2428 ± 0.1195 |

## Section 6. Parity Plots for Cluster-Specific and General Training Models

This section presents parity plots for individual cross-validation folds comparing cluster-specific and general training approaches for Cluster 1-1 and Cluster 1-2 interactions. Each plot displays predicted versus actual reactivity ratio values ($r_1$ and $r_2$) for a single fold.

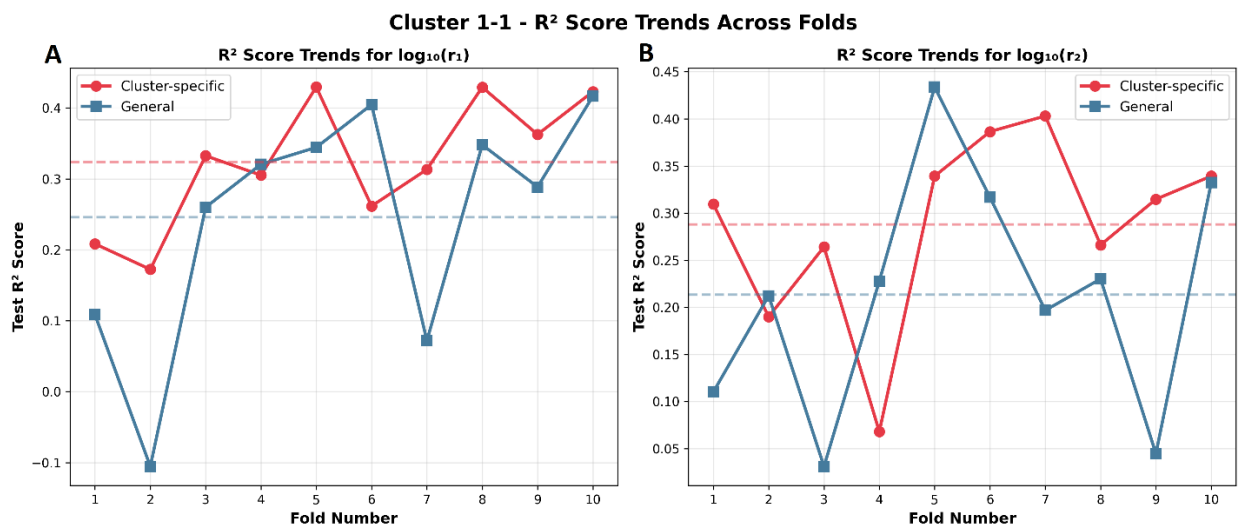

Figure S4. $R^2$ score trends across folds for Cluster 1-1 interaction: (A) $\log_{10}(r_1)$ and (B) $\log_{10}(r_2)$. Cluster-specific (red) vs. general (blue) training. Dashed lines show mean values.

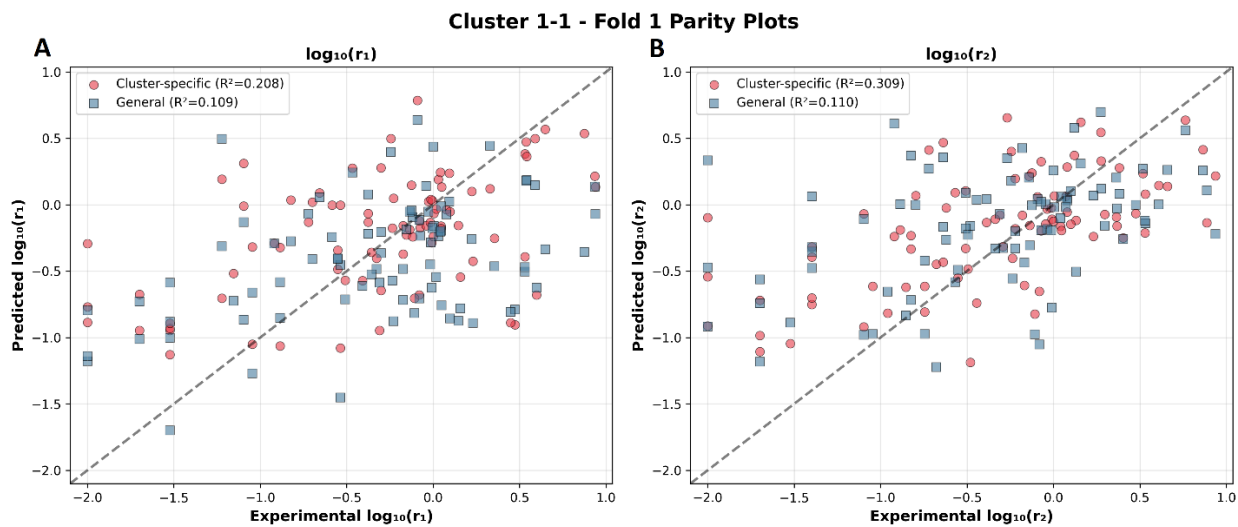

Figure S5. Parity plots for Cluster 1-1 interaction in Fold 1: (A) $\log_{10}(r_1)$ and (B) $\log_{10}(r_2)$. Cluster-specific (red circles) vs. general (blue squares) predictions. Dashed line represents perfect prediction.

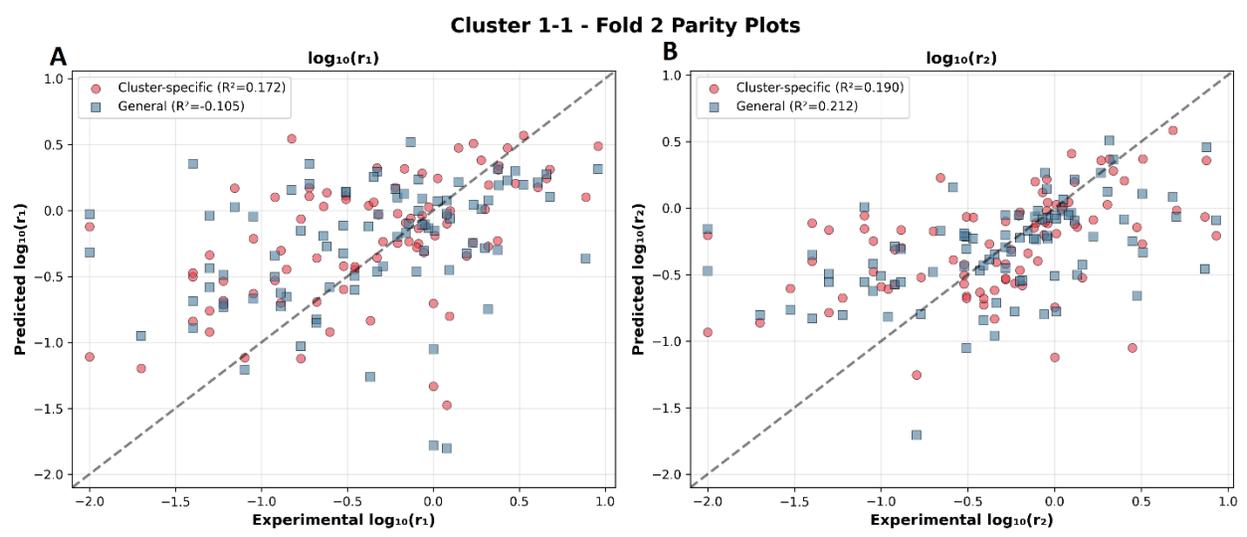

Figure S6. Parity plots for Cluster 1-1 interaction in Fold 2: (A) $\log_{10}(r_1)$ and (B) $\log_{10}(r_2)$. Cluster-specific (red circles) vs. general (blue squares) predictions. Dashed line represents perfect prediction.

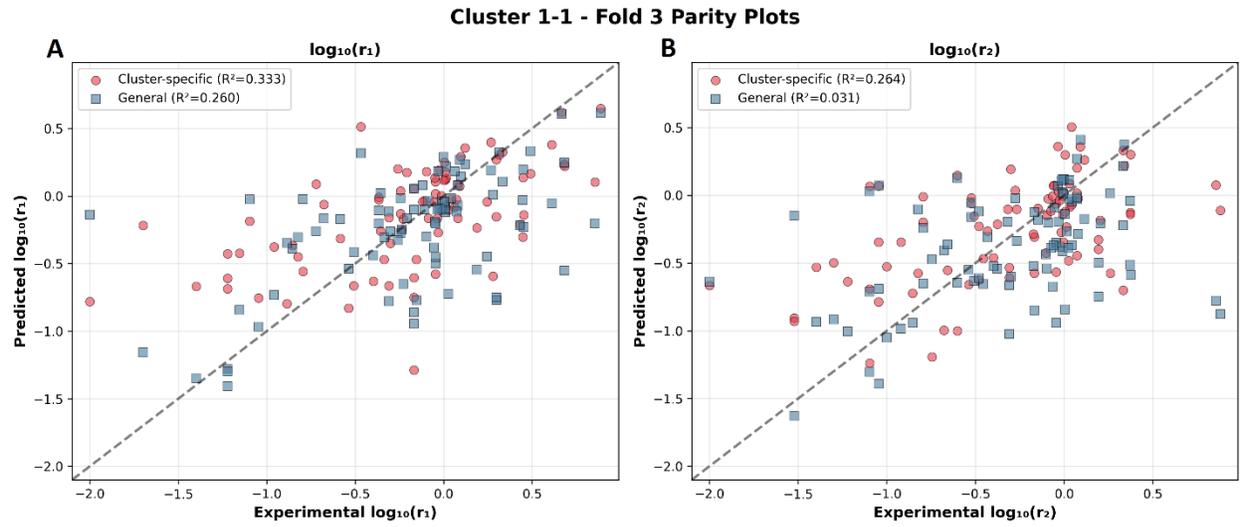

Figure S7. Parity plots for Cluster 1-1 interaction in Fold 3: (A) $\log_{10}(r_1)$ and (B) $\log_{10}(r_2)$. Cluster-specific (red circles) vs. general (blue squares) predictions. Dashed line represents perfect prediction

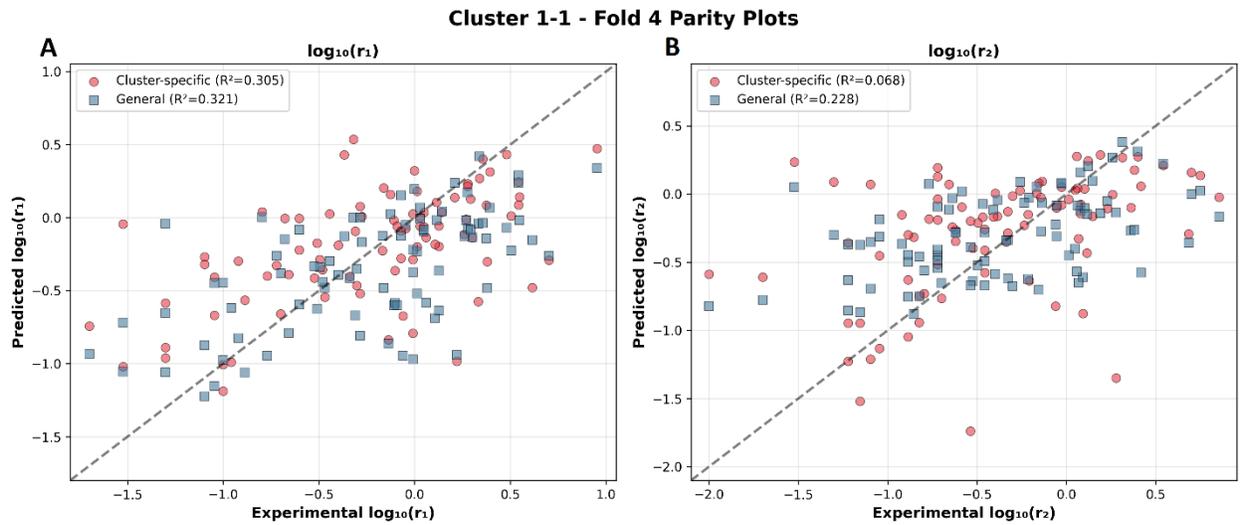

Figure S8. Parity plots for Cluster 1-1 interaction in Fold 4: (A) $\log_{10}(r_1)$ and (B) $\log_{10}(r_2)$. Cluster-specific (red circles) vs. general (blue squares) predictions. Dashed line represents perfect prediction

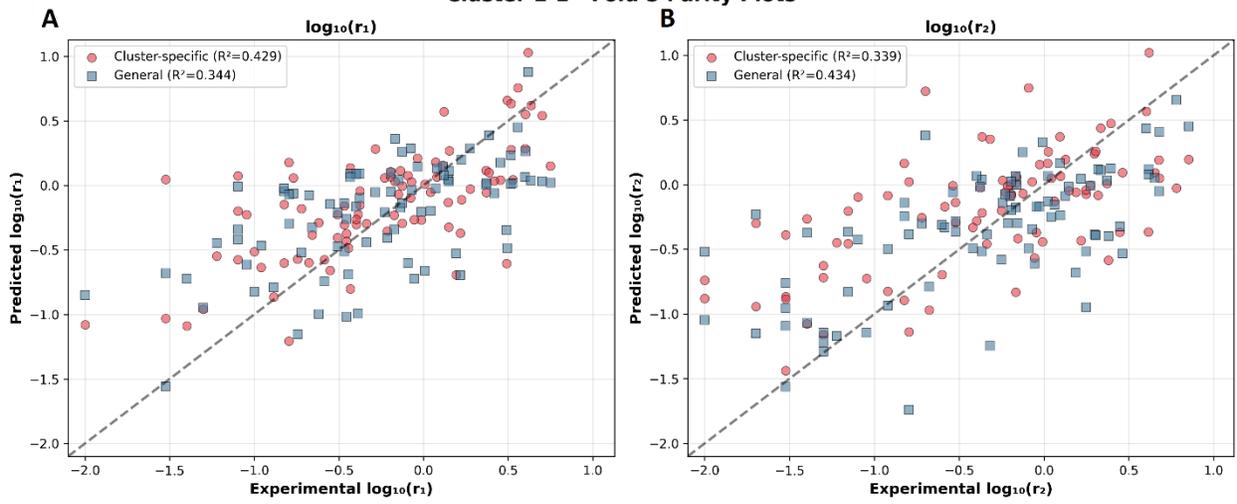

Figure S9. Parity plots for Cluster 1-1 interaction in Fold 5: (A) $\log_{10}(r_1)$ and (B) $\log_{10}(r_2)$. Cluster-specific (red circles) vs. general (blue squares) predictions. Dashed line represents perfect prediction

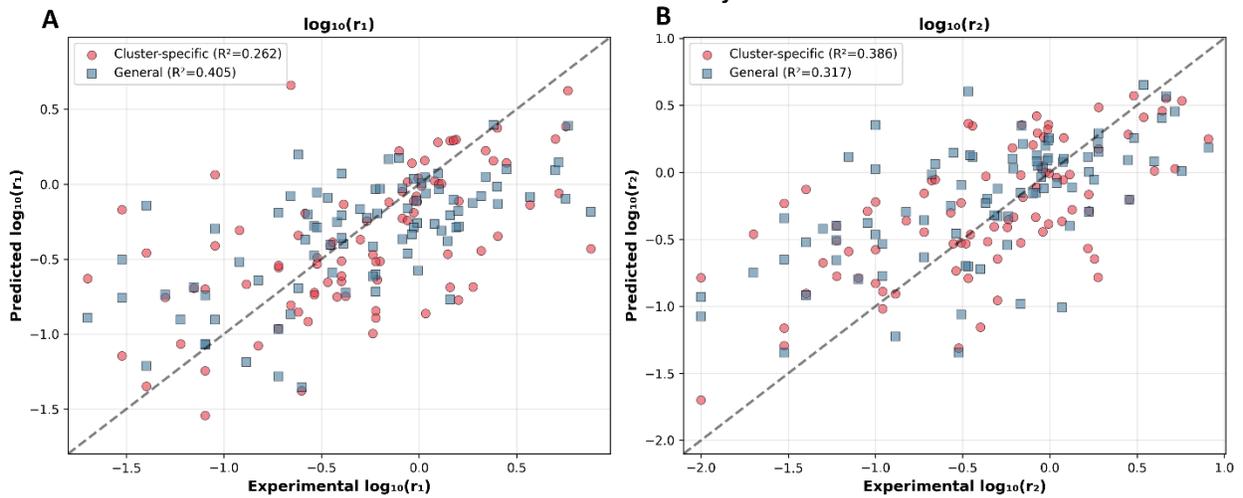

Figure S10. Parity plots for Cluster 1-1 interaction in Fold 6: (A) $\log_{10}(r_1)$ and (B) $\log_{10}(r_2)$. Cluster-specific (red circles) vs. general (blue squares) predictions. Dashed line represents perfect prediction.

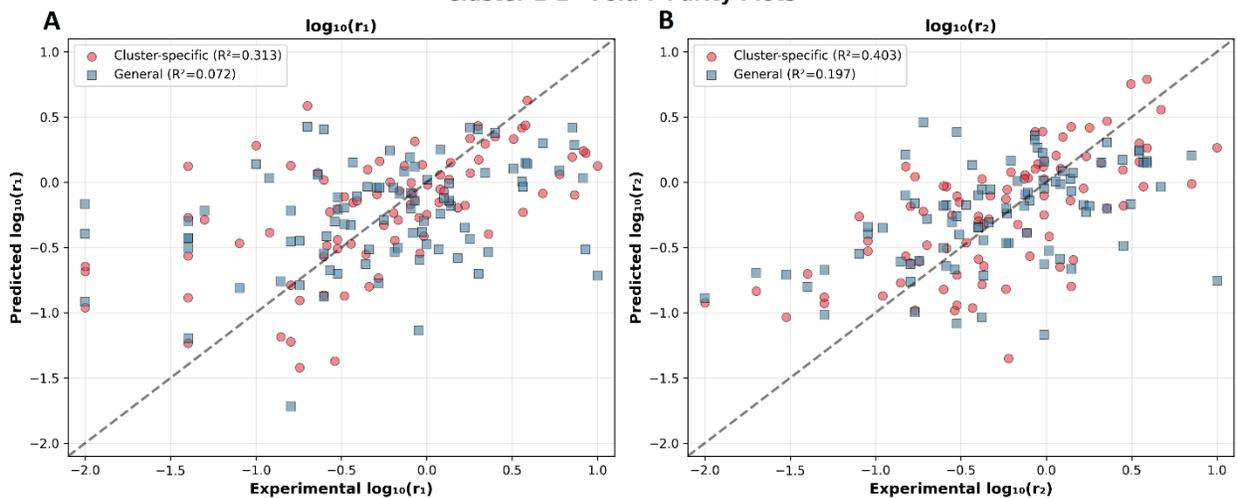

Figure S11. Parity plots for Cluster 1-1 interaction in Fold 7: (A) $\log_{10}(r_1)$ and (B) $\log_{10}(r_2)$. Cluster-specific (red circles) vs. general (blue squares) predictions. Dashed line represents perfect prediction.

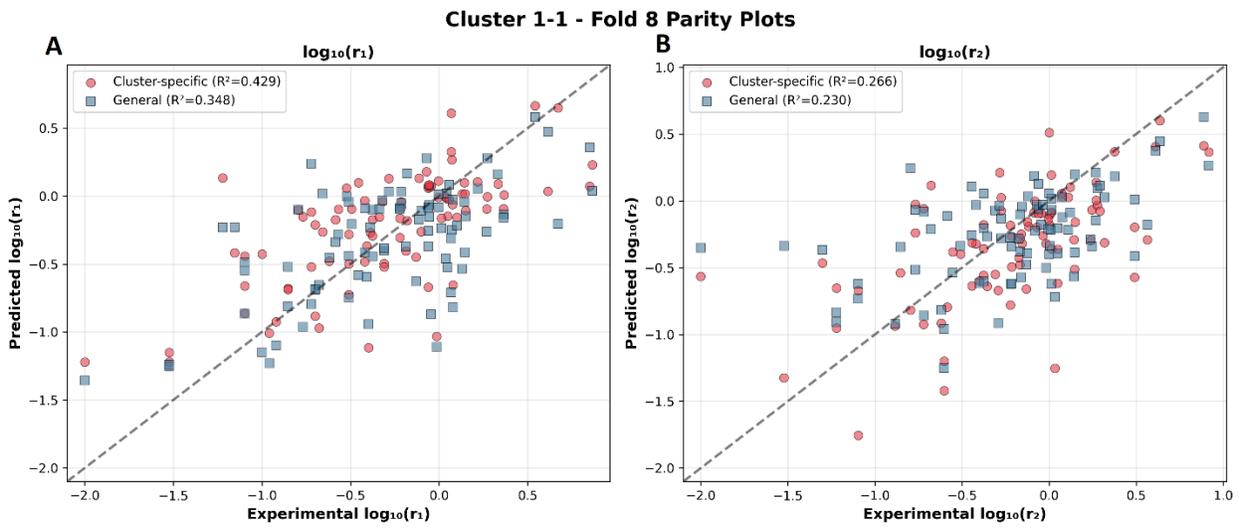

Figure S12. Parity plots for Cluster 1-1 interaction in Fold 8: (A) $\log_{10}(r_1)$ and (B) $\log_{10}(r_2)$. Cluster-specific (red circles) vs. general (blue squares) predictions. Dashed line represents perfect prediction.

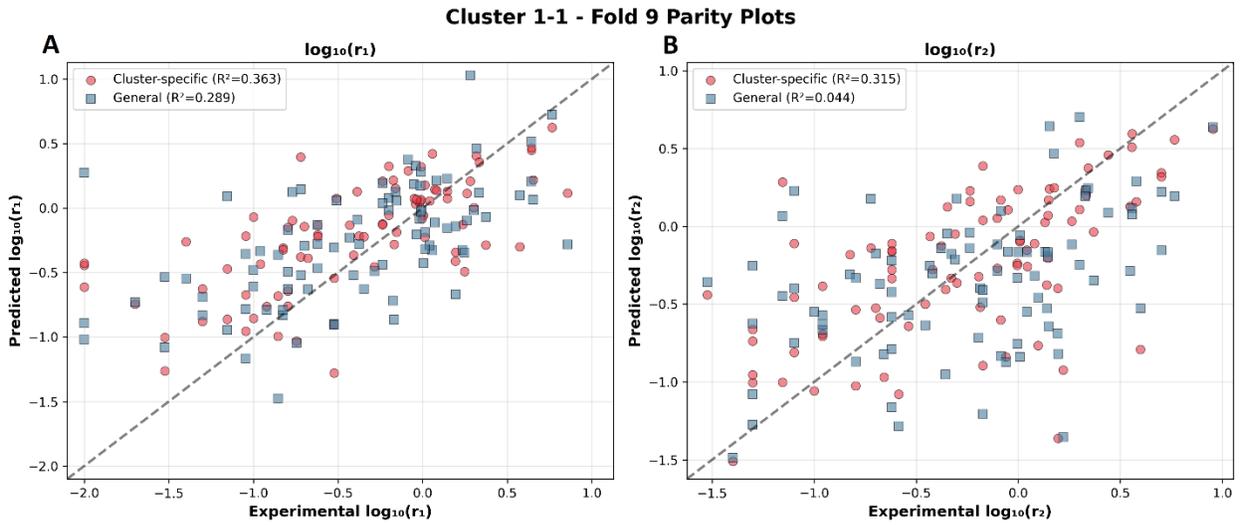

Figure S13. Parity plots for Cluster 1-1 interaction in Fold 9: (A) $\log_{10}(r_1)$ and (B) $\log_{10}(r_2)$. Cluster-specific (red circles) vs. general (blue squares) predictions. Dashed line represents perfect prediction.

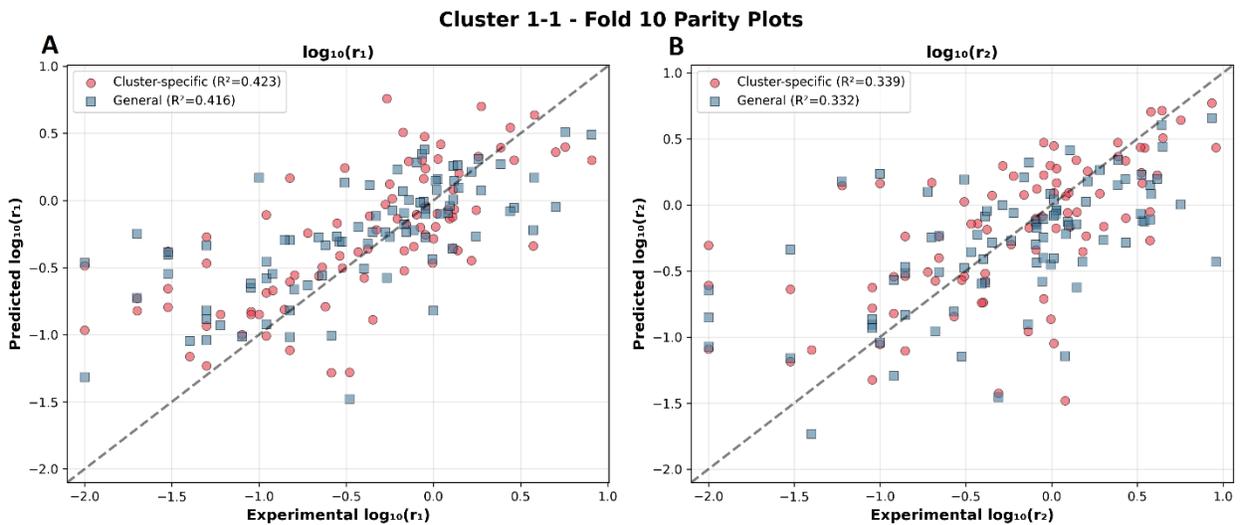

Figure S14. Parity plots for Cluster 1-1 interaction in Fold 10: (A) $\log_{10}(r_1)$ and (B) $\log_{10}(r_2)$. Cluster-specific (red circles) vs. general (blue squares) predictions. Dashed line represents perfect prediction.

Table S3. Performance metrics for Cluster-specific (Cluster 1-Cluster 1 interaction) vs. general predictions.

| Model | Train Mean Square Error (Log$_{10}$ r$_1$) | Train Mean Square Error (Log$_{10}$ r$_2$) | Test Mean Square Error (Log$_{10}$ r$_1$) | Test Mean Square Error (Log$_{10}$ r$_2$) | Train R$^2$ Score (Log$_{10}$ r$_1$) | Train R$^2$ Score (Log$_{10}$ r$_2$) | Test R$^2$ Score (Log$_{10}$ r$_1$) | Test R$^2$ Score (Log$_{10}$ r$_2$) | Test Mean Square Error (r$_1$) | Test Mean Square Error (r$_2$) | Test R$^2$ Score (r$_1$) | Test R$^2$ Score (r$_2$) |
|---|---|---|---|---|---|---|---|---|---|---|---|---|
| General Training | 0.0357 ± 0.0236 | 0.0369 ± 0.0174 | 0.2845 ± 0.0964 | 0.2880 ± 0.0545 | 0.9057 ± 0.0580 | 0.8999 ± 0.0492 | 0.2458 ± 0.1596 | 0.2135 ± 0.1211 | 2.0328 ± 0.9534 | 1.8981 ± 0.5452 | 0.0897 ± 0.1827 | 0.1271 ± 0.1664 |
| Cluster-Specific Training (Cluster 1-1 Interaction) | 0.0459 ± 0.0158 | 0.0440 ± 0.0111 | 0.2525 ± 0.0629 | 0.2626 ± 0.0510 | 0.8773 ± 0.0433 | 0.8823 ± 0.0304 | 0.3238 ± 0.0862 | 0.2880 ± 0.0941 | 1.7840 ± 0.7611 | 1.7348 ± 0.5425 | 0.1933 ± 0.0967 | 0.1890 ± 0.2265 |

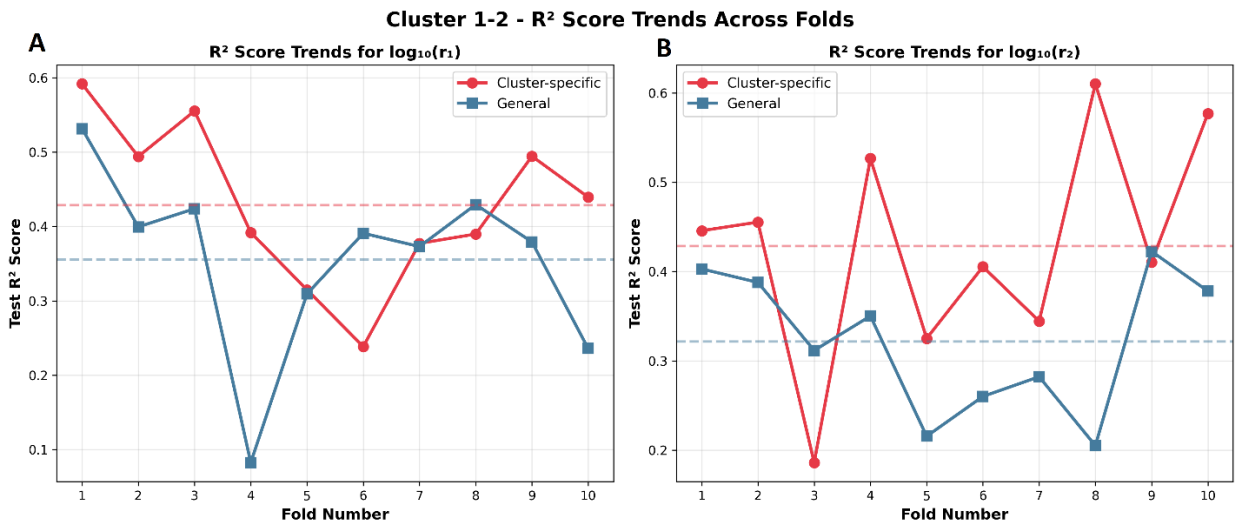

Figure S15. $R^2$ score trends across folds for Cluster 1-2 interaction: (A) $\log_{10}(r_1)$ and (B) $\log_{10}(r_2)$. Cluster-specific (red) vs. general (blue) training. Dashed lines show mean values.

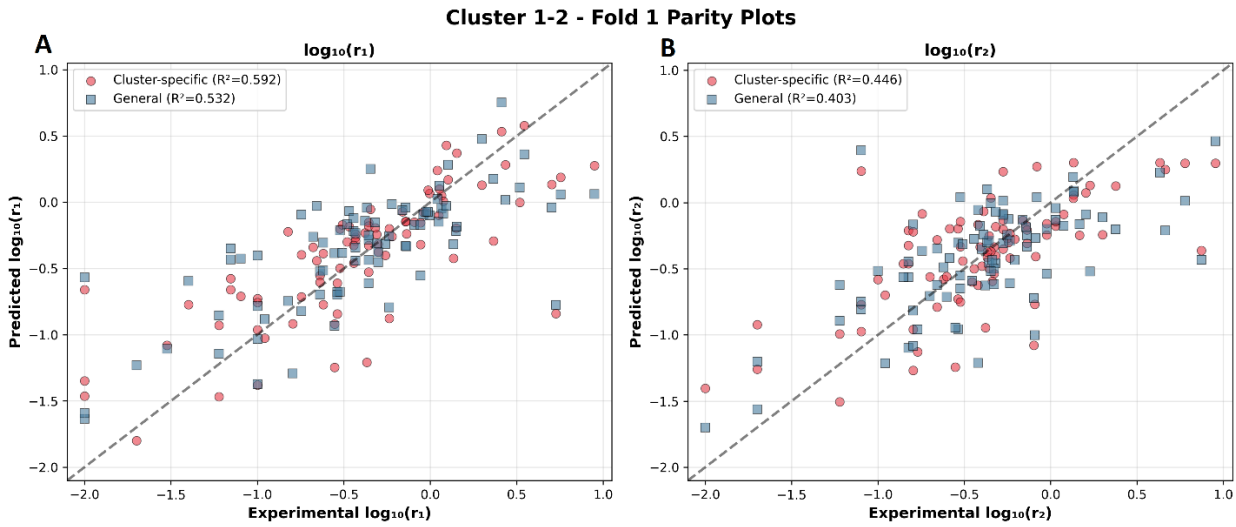

Figure S16. Parity plots for Cluster 1-2 interaction in Fold 1: (A) $\log_{10}(r_1)$ and (B) $\log_{10}(r_2)$. Cluster-specific (red circles) vs. general (blue squares) predictions. Dashed line represents perfect prediction.

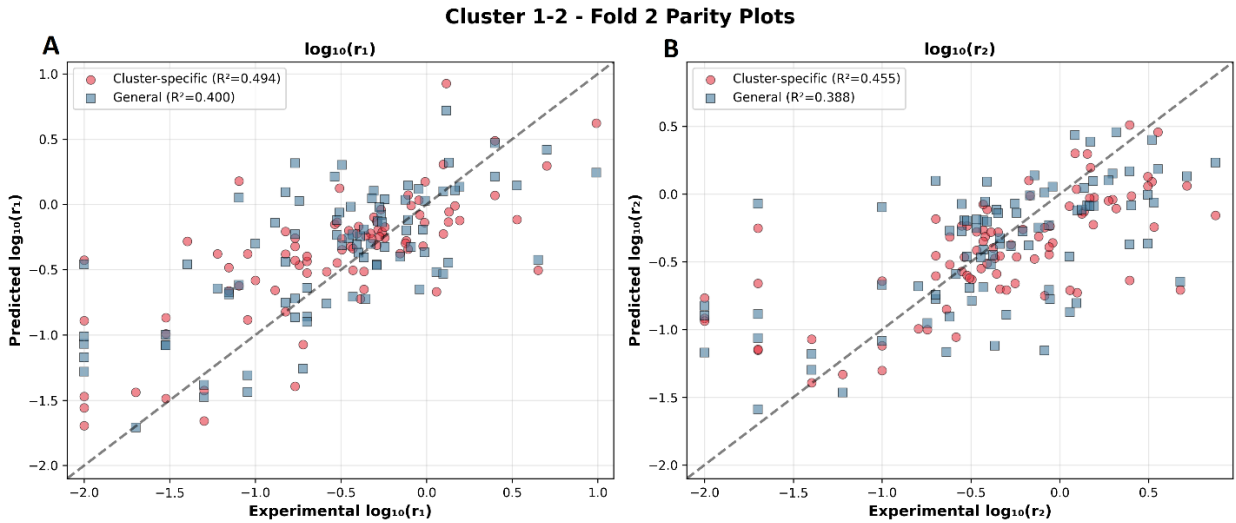

Figure S17. Parity plots for Cluster 1-2 interaction in Fold 2: (A) $\log_{10}(r_1)$ and (B) $\log_{10}(r_2)$. Cluster-specific (red circles) vs. general (blue squares) predictions. Dashed line represents perfect prediction.

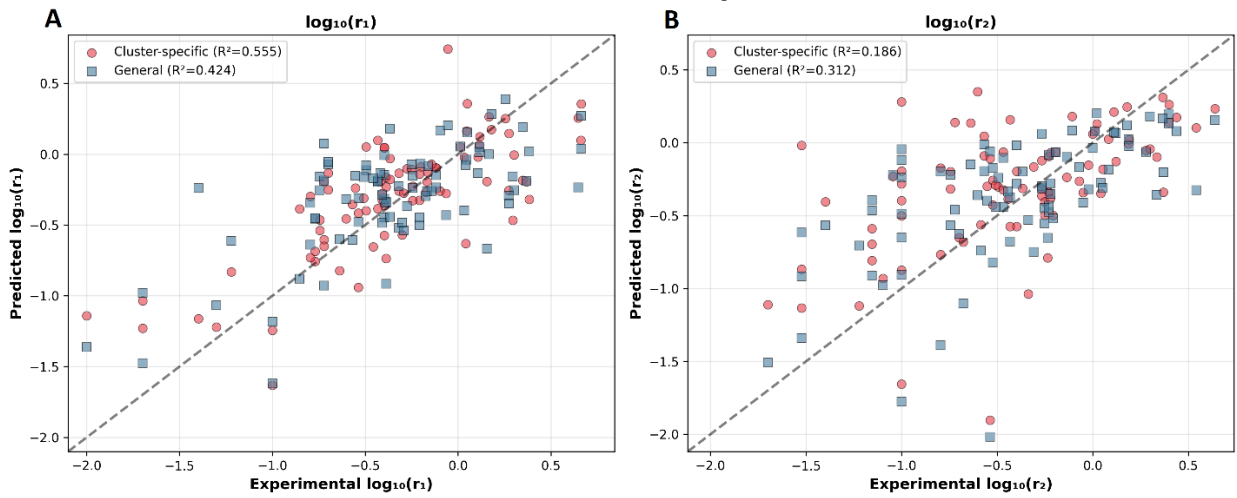

Figure S18. Parity plots for Cluster 1-2 interaction in Fold 3: (A) $\log_{10}(r_1)$ and (B) $\log_{10}(r_2)$. Cluster-specific (red circles) vs. general (blue squares) predictions. Dashed line represents perfect prediction.

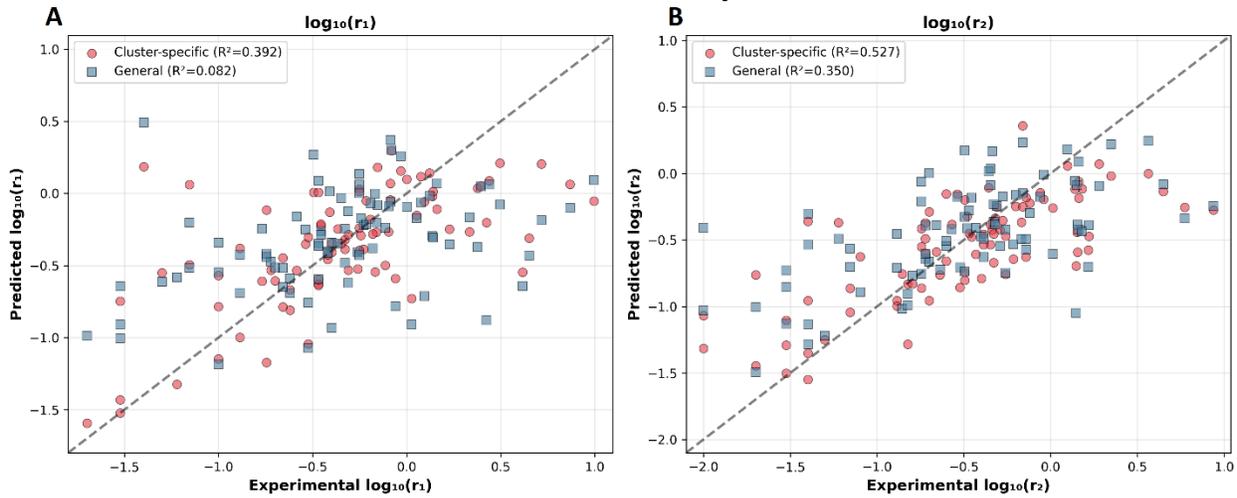

Figure S19. Parity plots for Cluster 1-2 interaction in Fold 4: (A) $\log_{10}(r_1)$ and (B) $\log_{10}(r_2)$. Cluster-specific (red circles) vs. general (blue squares) predictions. Dashed line represents perfect prediction.

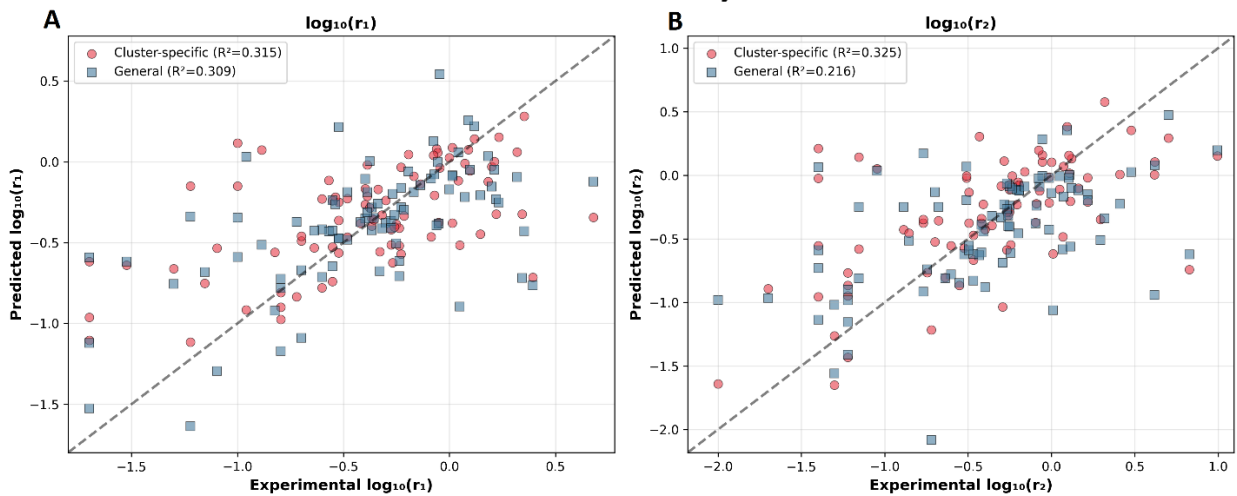

Figure S20. Parity plots for Cluster 1-2 interaction in Fold 5: (A) $\log_{10}(r_1)$ and (B) $\log_{10}(r_2)$. Cluster-specific (red circles) vs. general (blue squares) predictions. Dashed line represents perfect prediction.

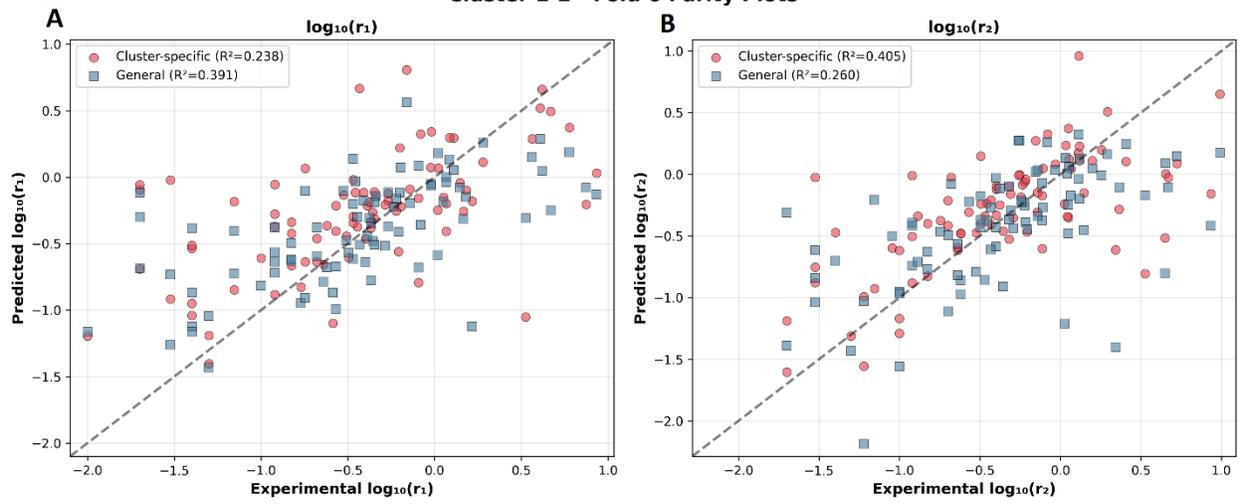

Figure S21. Parity plots for Cluster 1-2 interaction in Fold 6: (A) $\log_{10}(r_1)$ and (B) $\log_{10}(r_2)$. Cluster-specific (red circles) vs. general (blue squares) predictions. Dashed line represents perfect prediction.

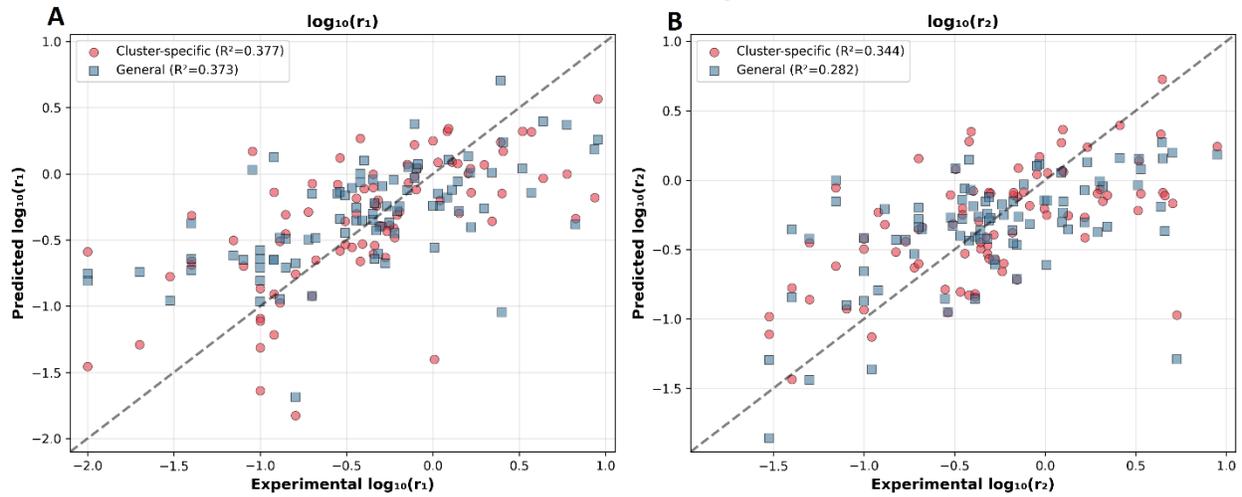

Figure S22. Parity plots for Cluster 1-2 interaction in Fold 7: (A) $\log_{10}(r_1)$ and (B) $\log_{10}(r_2)$. Cluster-specific (red circles) vs. general (blue squares) predictions. Dashed line represents perfect prediction.

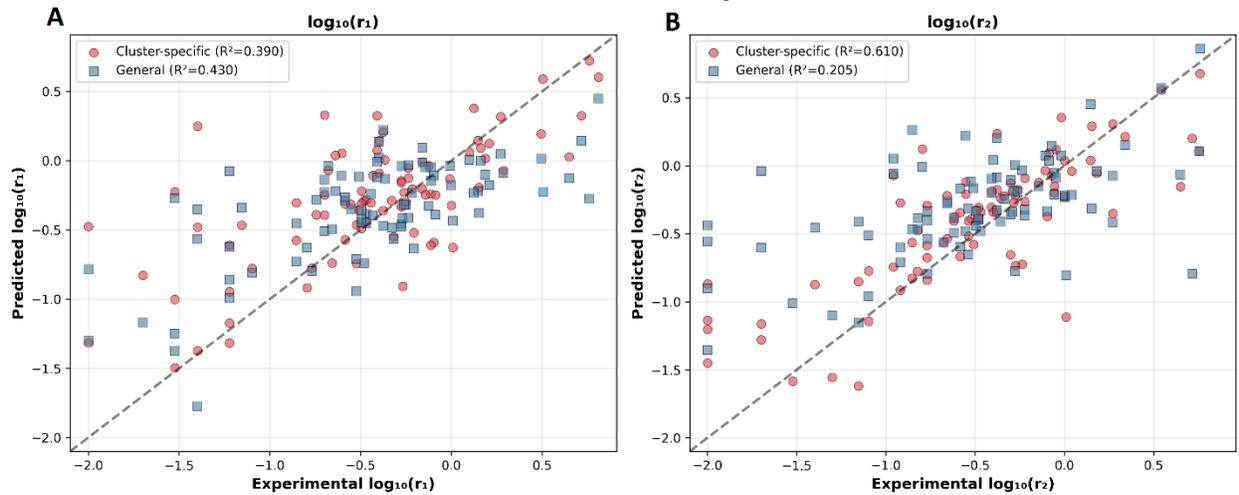

Figure S23. Parity plots for Cluster 1-2 interaction in Fold 8: (A) $\log_{10}(r_1)$ and (B) $\log_{10}(r_2)$. Cluster-specific (red circles) vs. general (blue squares) predictions. Dashed line represents perfect prediction.

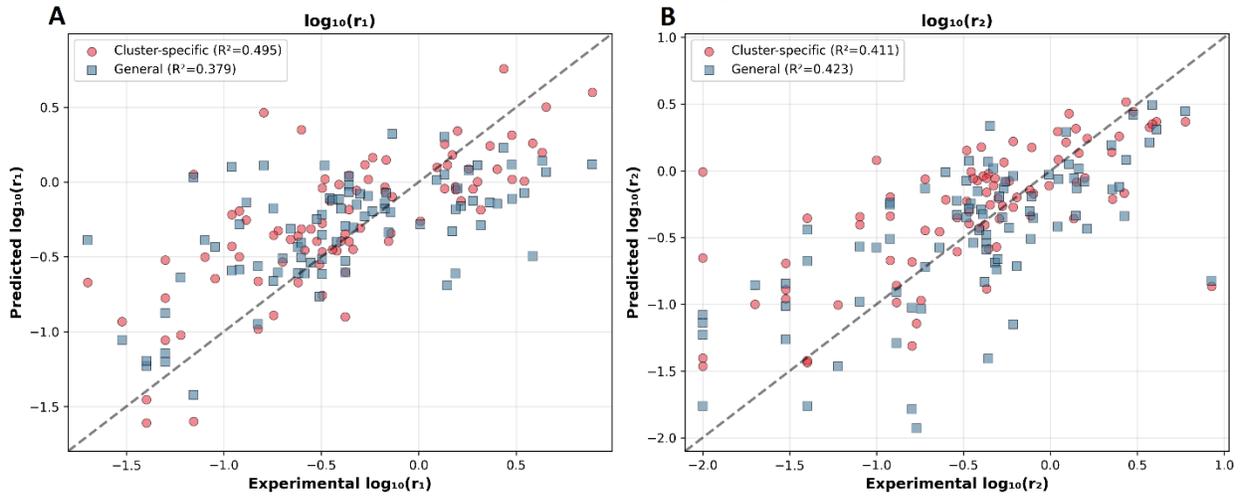

Figure S24. Parity plots for Cluster 1-2 interaction in Fold 9: (A) $\log_{10}(r_1)$ and (B) $\log_{10}(r_2)$. Cluster-specific (red circles) vs. general (blue squares) predictions. Dashed line represents perfect prediction.

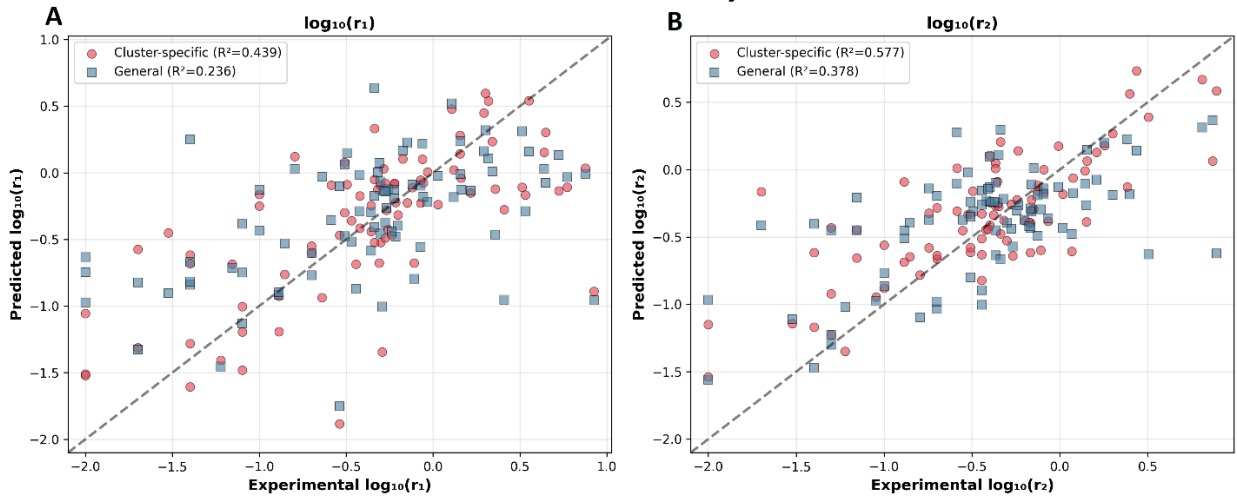

Figure S25. Parity plots for Cluster 1-2 interaction in Fold 10: (A) $\log_{10}(r_1)$ and (B) $\log_{10}(r_2)$. Cluster-specific (red circles) vs. general (blue squares) predictions. Dashed line represents perfect prediction.

Table S4. Performance metrics for Cluster-specific (Cluster 1-Cluster 2 interaction) vs. general predictions.

| Model | Train Mean Square Error (Log$_{10}$ r$_1$) | Train Mean Square Error (Log$_{10}$ r$_2$) | Test Mean Square Error (Log$_{10}$ r$_1$) | Test Mean Square Error (Log$_{10}$ r$_2$) | Train R$^2$ Score (Log$_{10}$ r$_1$) | Train R$^2$ Score (Log$_{10}$ r$_2$) | Test R$^2$ Score (Log$_{10}$ r$_1$) | Test R$^2$ Score (Log$_{10}$ r$_2$) | Test Mean Square Error (r$_1$) | Test Mean Square Error (r$_2$) | Test R$^2$ Score (r$_1$) | Test R$^2$ Score (r$_2$) |
|---|---|---|---|---|---|---|---|---|---|---|---|---|
| General Training | 0.0484 ± 0.0327 | 0.0503 ± 0.0350 | 0.2221 ± 0.0562 | 0.2333 ± 0.0365 | 0.8674 ± 0.0884 | 0.8623 ± 0.0942 | 0.3555 ± 0.1167 | 0.3216 ± 0.0745 | 1.6910 ± 0.7370 | 1.5603 ± 0.5686 | 0.0787 ± 0.1214 | 0.1703 ± 0.0943 |
| Cluster-Specific Training (Cluster 1-2 Interaction) | 0.0374 ± 0.0117 | 0.0377 ± 0.0144 | 0.1976 ± 0.0521 | 0.1947 ± 0.0382 | 0.8929 ± 0.0322 | 0.8914 ± 0.0425 | 0.4288 ± 0.1030 | 0.4286 ± 0.1196 | 1.5031 ± 0.7490 | 1.4952 ± 0.6609 | 0.1917 ± 0.1784 | 0.2088 ± 0.1708 |